\documentclass[usenatbib,usegraphicx,fleqn]{mn2e}
\usepackage{hyperref,amsmath,amssymb,xspace,times}

\newcommand{\SMILE}{\textsc{smile}\xspace}
\newcommand{\Nbody}{$N$-body\xspace}

\title[Schwarzschild method for disc galaxies]
{Applying Schwarzschild's orbit superposition method to barred \protect\\ or non-barred disc galaxies}

\author[E.~Vasiliev, E.~Athanassoula]
{Eugene Vasiliev$^1$\thanks{E-mail: eugvas@lpi.ru (EV), lia@oamp.fr (EA)}, E. Athanassoula$^{2}$\\
$^{1}$Lebedev Physical Institute, Leninsky prospekt 53, 119991, Moscow, Russia\\
$^{2}$Laboratoire d'Astrophysique de Marseille (LAM), UMR6110, CNRS/Universit\'e de Provence, 
38 rue Joliot Curie, 13388 \\ Marseille C\'edex 13, France}

\begin{document}

\date{Accepted 2015 April 9. Received 2015 April 9; in original form 2015 February 11}
\pagerange{2842--2856}\volume{450}\pubyear{2015}
\setcounter{page}{2842}

\maketitle


\begin{abstract}
We present an implementation of the Schwarzschild orbit superposition method 
which can be used for constructing self-consistent equilibrium models of barred or
non-barred disc galaxies, or of elliptical galaxies with figure rotation.
This is a further development of the publicly available code \SMILE; its main improvements 
include a new efficient representation of an arbitrary gravitational potential using 
two-dimensional spline interpolation of Fourier coefficients in the meridional plane, 
as well as the ability to deal with rotation of the density profile and with multicomponent 
mass models. We compare several published methods for constructing composite axisymmetric 
disc--bulge--halo models and demonstrate that our code produces the models that are closest 
to equilibrium. We also apply it to create models of triaxial elliptical galaxies with 
cuspy density profiles and figure rotation, and find that such models can be found and are 
stable over many dynamical times in a wide range of pattern speeds and angular momenta, 
covering both slow- and fast-rotator classes. We then attempt to create models of strongly 
barred disc galaxies, using an analytic three-component potential, and find that it is not 
possible to make a stable dynamically self-consistent model for this density profile. 
Finally, we take snapshots of two \Nbody simulations of barred disc galaxies embedded in 
nearly-spherical haloes, and construct equilibrium models using only information on 
the density profile of the snapshots.
We demonstrate that such reconstructed models are in near-stationary state, in contrast with 
the original \Nbody simulations, one of which displayed significant secular evolution.
\end{abstract}


\section{Introduction}

An important task that often arises in stellar dynamics is the construction of self-consistent 
equilibrium models of galaxies that have a given density profile $\rho(\boldsymbol{x})$ and, 
optionally, satisfy certain kinematic constraints. 
Self-consistency means that the gravitational potential is determined by the distribution of 
the mass in the galaxy via the Poisson equation. 
Equilibrium models satisfy the collisionless Boltzmann equation with a steady-state 
distribution function. In practice, we usually require the models to evolve very slowly 
compared to the dynamical (crossing) timescale.

There exist numerous methods for creating (near-)equilibrium galaxy models of varying complexity 
and geometry. After reviewing of these approaches in Section~\ref{sec:overview}, 
with a particular focus on the methods suitable for disc galaxies, we concentrate on 
the Schwarzschild orbit superposition method. 
Up to now, it was mostly applied to elliptical galaxies or to the inner parts of the Milky Way. 
We describe the challenges that need to be met in its extension to highly flattened, 
non-axisymmetric stellar systems with figure rotation, and present our implementation in 
Section~\ref{sec:method_details}. 
Then in Section~\ref{sec:tests} we apply it to construct models of some analytical 
density profiles that are frequently used in the literature, in particular, comparing several 
published methods for creating equilibrium models of axisymmetric disc galaxies, and create 
models of barred disc galaxies based on the density distributions taken from \Nbody simulations.
We discuss our results and possible future applications in Section~\ref{sec:discussion},
and summarize in Section~\ref{sec:conclusions}.

\section{Methods for creating equilibrium models}  \label{sec:overview}

The focus of this paper is on disc galaxies, therefore we restrict our review to 
the methods suitable for such systems which are typically multicomponent 
(include a strongly flattened disc, which may or may not be axisymmetric, a rather spherical 
halo component, and a bulge or a bar that might well be triaxial).

Most of the methods specifically tailored for disc galaxies produce axisymmetric models that 
are close but not exactly in equilibrium.
\citet{Hernquist1993} introduced a method based on the locally Maxwellian approximation to 
the velocity distribution function and a prescribed density distribution for each component. 
It used the spherical (for the halo) and axisymmetric (for the bulge) Jeans equations and 
an approximation of constant radial-to-vertical velocity dispersion ratio for the disc 
together with epicyclic approximation for the azimuthal component of velocity dispersion.
In this method, all components are spherical except, in some parts of the analysis, the disc;
later \citet{BoilyKP2001} generalized this method to allow for axisymmetric halo and bulge, 
while using an approximate method to squeeze the shape of velocity ellipsoid proportionally 
to the flattening of the potential. 
A drawback of this approach is the use of Maxwellian velocity distribution function, which may 
not always be appropriate for an equilibrium solution \citep{KazantzidisMM2004}.

\citet{KuijkenDubinski1995} used a different approach, starting from a suitable anzatz for 
the distribution functions of each component, which used two classical integrals -- energy and 
$z$-component of angular momentum for the spheroidal components, and added the third approximate 
integral (energy of vertical motion) for the disc. The density profile corresponding to 
the given distribution function was then computed iteratively, taking into account the potentials 
of all components. \citet{WidrowDubinski2005} modified this scheme to allow for a cuspy density 
profile and a central massive black hole, while retaining analytic expressions for 
the distribution functions. However, with this approach the density profile of the halo and bulge 
are determined implicitly and are hard to prescribe; an improvement introduced by 
\citet{WidrowPD2008} instead derived the isotropic distribution function for the given density 
profile numerically, using the Eddington inversion formula in a spherically-symmetric 
approximation of the total potential. A similar approach (iterative solution for the potential 
for given analytically described distribution functions for each component) is employed 
in the widely used Besan\c con model of Milky Way \citep{RobinRDP2003}.
Binney and collaborators advocate the use of actions as arguments of distribution function 
\citep[e.g.][]{BinneyMcMillan2011}; however, their models are not designed to be self-consistent 
dynamically and instead focus on the fitting of kinematic properties of observed stellar 
populations in the Milky Way disc.

Another possible approach is to build up the composite system by introducing the components 
of gravitational potential one by one, in the adiabatic approximation. 
In the first such implementation by \citet{Barnes1988}, one begins with \Nbody realizations 
of two separate analytic spherical equilibrium models for bulge and halo, superimposing them 
on top of each other and allowing them to relax.
This first stage, however, induces non-negligible changes to the original density profiles as 
they reach the new equilibrium in the joint gravitational potential.
Then the analytically defined disc potential is grown adiabatically, without actually 
adding any disc particles, but allowing the particles in the spheroidal components to adjust 
to the combined potential.
Finally, the disc is populated with particles using the locally Maxwellian distribution function. 
\citet{McMillanDehnen2007} improve this method by constructing the spheroidal  
components from an exact distribution function derived from an anisotropic generalization of 
Eddington inversion formula, using the monopole terms of potential of all three components. 
Then only the non-spherical part of the disc potential is adiabatically introduced into the live 
\Nbody system of bulge plus halo, and finally the disc particles are populated from a more 
accurate distribution function, avoiding the Maxwellian approximation. This method offers 
little control over the shape of the halo or bulge (which slightly flatten during adiabatic 
contraction), but yields close-to-equilibrium configurations even for rather ``warm'' discs.

All of the above mentioned methods are designed for axisymmetric potentials only, 
often make an implicit assumption that the motion of stars in the disc component is integrable
\citep[e.g.\ in the torus construction method of][]{BinneyMcMillan2011}, and usually have 
a rather restricted choice for the mass and velocity profiles of various components.
It has long been recognized that non-axisymmetric features in galactic discs, such as 
bars, spiral arms and rings, give rise to a variety of phenomena associated with resonances 
and chaotic orbits \citep[e.g.][]{Dehnen2000a,Fux2001,RomeroGomez2006}.
Many studies have been devoted to the detailed analysis of orbital structure of barred disc 
galaxies described by analytical potentials 
\citep[e.g.][]{ContopoulosP1980,AthanassoulaBMP1983,Pfenniger1984a,TeubenSanders1985,SkokosPA2002} 
or \Nbody models \citep[e.g.][]{PfennigerFriedli1991,Berentzen1998,HarsoulaKalapotharakos2009,
Fragkoudi2015}. To address the question which of the many kinds of orbits actually contribute to 
the self-consistent potential, so-called response models are constructed, following the method 
developed by \citet{ContopoulosGrosbol1988}. It starts from an assumption for the gravitational 
potential (including non-axisymmetric perturbations and a value for pattern speed), 
then the periodic orbits in the given potential are computed and superimposed to obtain the 
``response density'' profile. The parameters of the potential are then iteratively adjusted 
until the response density matches the potential. A number of indicators are used to determine 
self-consistency, including the amplitude and phase of various angular harmonics as functions 
of radius. 
This technique has been applied to barred spiral galaxies \citep[e.g.][]{KaufmannContopoulos1996}; 
a modification introduced in \citet{KalapotharakosPG2010} allowed to vary the relative 
contribution of different groups of orbits (binned in energy) in the solution, similar to 
the Schwarzschild method (see below). 
An advantage of this method is that it uses the knowledge of the orbital structure of 
non-axisymmetric galaxies; nevertheless, it still is limited by the assumptions of 
the functional form of potential and perturbation, and has been only used in two dimensions.

A completely different and very generic alternative is offered by combining \Nbody simulations 
(which by definition are dynamically self-consistent, even though not necessarily stationary) 
with some adjustment method to bring the model towards the required properties.
The Made-to-Measure (M2M) method, introduced by \citet{SyerTremaine1996}, deals with a fixed, 
pre-determined potential, and adjusts the weights of individual particles in the course of 
simulation to match the prescribed observable properties. \citet{BissantzDG2004} and 
\citet{deLorenziDGS2007} extended this method to rotating potentials and included 
a likelihood-based treatment of kinematic observational constraints. This implementation also 
has an option of adjusting the gravitational potential together with the evolving density
(using a spherical-harmonic expansion), thus making the resulting model dynamically self-consistent. 
\citet{Dehnen2009} improved the method by introducing a new scheme for adjusting particle weights, 
which has later been incorporated into other codes.
\citet{LongMao2010} and \citet{HuntKawata2013} developed two more implementations of M2M 
method, applicable also to disc galaxies \citep{LongMSW2013,HuntKM2013}.
Another class of methods based on ``guided'' \Nbody simulations are the iterative methods 
of \citet{RodionovAS2009} and \citet{YurinSpringel2014}. In these approaches, the evolution 
of a live \Nbody system is followed for a certain time (an episode), and then the velocities 
of particles are adjusted according to some scheme, to match the required properties of the system. 
The whole process is repeated iteratively until a convergence is reached. 
Both M2M and iterative methods are very flexible and capable of dealing with any geometry 
(although the method of \citealt{YurinSpringel2014} is restricted to axisymmetry); they are however 
rather costly in terms of computational demand. 

Finally, the orbit superposition method, introduced by \citet{Schwarzschild1979}, 
has been widely used in the last decade to create stellar-dynamical models of external galaxies.
There exist several implementations for axisymmetric geometry \citep{CrettonZMR1999,Gebhardt2000,
ValluriME2004} and one for non-rotating triaxial potentials with cores \citep{vdBosch2008}, 
which are capable of dealing with a variety of observational constraints and perform a model 
search to find the best-fitting parameters of the potential and their uncertainties.
There are also implementations targeted for a more theoretical usage (without observational 
constraints), e.g.\ \citet{MerrittFridman1996,Siopis1998,Thakur2007,Vasiliev2013}.
All these studies considered galactic models that are not too far from spherical, 
i.e.\ they could be axisymmetric or triaxial, but not strongly flattened. 
Applications to barred disc galaxies are rather scarce: \citet{Pfenniger1984b} and 
\citet{WozniakPfenniger1997} considered two-dimensional models for the motion in the galactic 
plane of a barred potential, and several studies \citep{Zhao1996b, HafnerEDB2000, 
WangZMR2012, WangMLS2013} concentrated on the models of Milky Way bulge, without attempting 
to construct a model for the entire disc.

In this paper, we present a publicly available\footnote{\url{http://td.lpi.ru/~eugvas/smile/}}
implementation of the Schwarzschild method suitable to deal with multicomponent, arbitrarily 
flattened, non-axisymmetric potentials with figure rotation. 
It is a further development of the computer code \SMILE, described in detail in \citet{Vasiliev2013}.
In the present version, it remains ``a theorist's tool'', in the sense that it does not include 
the possibility to deal with observational kinematic constraints; however, it offers a very 
generic framework suitable for all types of galaxies, and in the future we plan an observational 
extension of the code. 

\section{Technical details of Schwarzschild modelling}  \label{sec:method_details}

To construct a self-consistent equilibrium model with the Schwarzschild method, one takes 
the following steps:
\begin{enumerate}
\item Choose a model for the gravitational potential.
\item Compute a large number (typically $\gtrsim 10^4$) of orbits in this potential, sampling, 
in as much as possible, the entire phase space. Each orbit is computed for many ($\sim 100$) 
dynamical times and its properties are stored in a discretized way (see Section~\ref{sec:solution}). 
\item The model is constructed as a weighted superposition of these structural blocks (orbits), 
with the weights being computed as a solution to some optimization problem. 
\end{enumerate}
If the method is used in the observational context, then a series of models with varied 
parameters of the potential is constructed, and the likelihood of each model in fitting 
the observational data is used to derive the confidence ranges of the potential parameters.

We describe these steps in detail in the following sections, focusing on the particular 
features that are necessary to deal efficiently with models of disc galaxies.

\subsection{Potential approximation}  \label{sec:potential}

At the very least, a flexible and accurate method for representing an arbitrary 
potential of a highly flattened system is required. 
Of course, one may use a combination of analytical potential-density models which approximate 
the target system, such as a Miyamoto--Nagai disc or a Ferrers triaxial bar, which have been 
used in many previous studies \citep[e.g.,][]{AthanassoulaBMP1983, Pfenniger1984a}.
This approach may well capture important properties of the system, but it is difficult to 
control the systematic errors from parametric models. 
On the other hand, approximating the potential by a suitably chosen basis-set expansion 
is a more general approach, although it is not trivial to do it efficiently for a strongly 
flattened system. 
Most work that has been done in this direction is based on the spherical-harmonic expansion 
of both density and potential up to a given order $l_\mathrm{max}$ in the angles, and 
expressing the radial dependence of expansion coefficients as a sum over basis functions 
\citep{CluttonBrock1973,HernquistOstriker1992,Zhao1996a,Weinberg1999}, 
or as explicit functions of radius \citep{McGlynn1984,Sellwood2003,Vasiliev2013}. 
This technique, especially in its last variant, offers a very good approximation of the potential 
of systems which are not too far from spherical (i.e.\ for elliptical galaxies), but becomes 
increasingly inefficient for disky systems as the order of angular expansion, necessary to 
resolve the thinness of the disc, increases (for instance, \citet{HolleyWK2005} used 
$l_\mathrm{max}=36$).

On the other hand, for strongly flattened systems it is more natural to work in cylindrical
coordinates $R,z,\phi$ rather than spherical $r,\theta,\phi$.
For two-dimensional systems, an efficient basis set may be constructed using appropriately 
scaled Bessel functions \citep{CluttonBrock1972,Qian1993}. 
The third dimension may be added in several different ways. 
\citet{Earn1996} presents a basis set for density models separable in $R,z$, 
while \citet{RobijnEarn1996} generalize this approach for other coordinate systems besides 
cylindrical. \citet{Weinberg1999} resorts to a numerical solution of the Sturm--Liouville 
equation to obtain the basis functions in $R$, while using harmonic expansion in $\phi$ 
and $z$ (the latter in a finite-size sheet). \citet{BrownPapaloizou1998} employ a similar 
approach with both $R$ and $z$ basis functions being expressed by suitably transformed 
polynomials, tailored to a given lowest-order density distribution. 
In a special case of a separable axisymmetric density profile in cylindrical coordinates, 
\citet{KuijkenDubinski1995} and \citet{DehnenBinney1998} used an approximation based on  
the splitting of the potential into a separable exactly representable flattened part and 
a spherical-harmonic expansion for the remaining weakly non-spherical part. 
Finally, in \Nbody simulations often a three-dimensional cylindrical grid is used 
\citep[e.g.][]{PfennigerFriedli1993,SellwoodValluri1997}; however, it usually employs 
a low-order (linear) interpolation for force computation, which might be insufficient for 
high-accuracy orbit integration.

Given a good experience of working with a non-parametric representation of spherical-harmonic 
expansion coefficients as spline functions in scaled radius, 
we decided to use a similar approach for flattened systems in cylindrical coordinates.
Namely, we represent the potential as a sum of Fourier terms in the azimuthal angle $\phi$, 
with the coefficients of expansion being smooth functions in the meridional plane $R,z$, 
that is, two-dimensional cubic splines in suitably scaled coordinates. 
We refer to the appendix for a more technical description and accuracy tests of this potential 
approximation. Overall, its performance is good enough to be used for approximating an 
arbitrary flattened non-axisymmetric mass distribution. Moreover, in multicomponent models 
we may choose the most suitable potential representation for each component, for instance, 
the cylindrical spline for the disc galaxy and the spherical-harmonic spline expansion for 
the halo.

\subsection{Generation of initial conditions for the orbit library}  \label{sec:initial_conditions}

The core feature of the Schwarzschild method is that it automatically picks up the orbits 
that are most suitable for representing the target density profile. Nevertheless, the efficiency 
of the method and the very existence of the solution do depend on the procedure employed to sample 
the initial conditions of the orbit library. 
As an extreme example, suppose we wish to create a spherical system in dynamical equilibrium 
using purely radial orbits (even though it is guaranteed to undergo a violent radial-orbit 
instability, this does not invalidate the concept of equilibrium solution). 
If we pick up initial velocities with random orientations, for instance using the isotropic 
distribution function, the chances to find a purely radial orbit are zero. 
On the other hand, if we use a spherical Jeans equation with the velocity anisotropy coefficient 
$\beta \equiv 1-\sigma_\mathrm{t}^2/(2\sigma_\mathrm{r}^2)$ equal to unity (where 
$\sigma_\mathrm{r}$ and $\sigma_\mathrm{t}$ are the velocity dispersion in the radial and 
tangential directions, respectively), then all our orbits will be radial. 
We could also start all orbits with zero velocities, and do not even need to seed 
the initial positions of the orbits to follow the required density profile -- 
the solution obtained with the Schwarzschild method will converge to it, as long as we ensure 
that there are sufficiently many (not necessarily all) radial orbits in the orbit library.
As another example, using isotropic velocities in a rotationally-supported disc galaxy 
would be much less efficient than, for instance, launching orbits with some small velocity 
dispersion about the mean circular velocity; however, if this velocity dispersion is too small
for a given thickness of the disc, a self-consistent model could not be constructed either.

This suggests that the choice of initial conditions is an important ingredient of the model, 
and should, in as much as possible, correspond to the target system to be constructed.
In the traditional approach, used in nearly all previous studies 
\citep[e.g.][]{MerrittFridman1996,vdBosch2008}, one takes a small number 
of discrete values of energy $E$ and populates the initial conditions from several 
``start-spaces'' at each energy. Stationary start-space \citep{Schwarzschild1993} consists of 
points covering the equipotential surface with zero initial velocities; principal-plane 
start-spaces contain points with one coordinate (say, $z$) and two complementary velocity 
components ($v_x,v_y$) being zero, and the remaining component $v_z$ assigned to yield 
the required total energy. 
In the axisymmetric case, it is sufficient to sample the positions in the meridional plane 
on the zero-velocity curve for the given values of energy and angular momentum 
\citep[e.g.][]{CrettonZMR1999}. For models with figure rotation, \citet{Schwarzschild1982} 
proposed a $y-\alpha$ start space, with initial conditions taken along the intermediate ($y$) 
axis and with velocity directed at the angle $\alpha$ in the $x-z$ plane.

All these methods place initial conditions on some sort of regular grid, which may not always 
be a good choice. For instance, with $10^4$ orbits we only have $\sim20$ grid nodes per each 
dimension (the grid of initial conditions for triaxial models is usually three-dimensional), 
which leads to excessive granularity of the orbit library. 
As shown by \citet{VasilievAthanassoula2012}, such models are not in perfect equilibrium and 
demonstrate a noticeable evolution in the course of \Nbody simulations at early times. 
An alternative approach, used in the aforementioned paper, is to assign initial conditions 
randomly, for instance, sampling them from an isotropic distribution function computed with 
the Eddington inversion formula \citep[e.g.][equation~4.46]{BinneyTremaine} for a spherical 
potential--density pair that approximates the real, non-spherical model. 
In this approach, the positions of particles are first seeded in accordance with the true 
three-dimensional density profile of the model, then the velocities are assigned by drawing 
them from the Eddington distribution function.

In the present implementation, we introduced several additional methods for generating 
the initial conditions for velocities, aside from the Eddington sampler. 
The first two are based on the Jeans equations for the spherical and axisymmetric systems. 
The former is rather trivial, and its only improvement over the Eddington sampler is 
the facilitation of creating models with strong radial or tangential velocity anisotropy%
\footnote{We do not argue that the Jeans equation approach is superior to the 
distribution-function-based method, and it is known that the former may lead to significant 
biases in some applications \citep{KazantzidisMM2004}. However, in the Schwarzschild method 
the distribution function is recovered numerically, and thus these drawbacks are irrelevant.}.
The axisymmetric Jeans equations are used in the anisotropic formulation of \citet{Cappellari2008}: 
the velocity dispersion ellipsoid is assumed to be aligned with the cylindrical coordinate system, 
and the anisotropy coefficient in the meridional plane 
$\beta_\mathrm{m} \equiv 1-\sigma_z^2/\sigma_R^2$ 
is assumed to be constant. These assumptions are rather strong and moreover they are not satisfied 
for realistic galaxies; for an alternative method of solving the axisymmetric Jeans equations 
under assumption of radial alignment of velocity ellipsoid see \citet{YurinSpringel2014}. 
Nevertheless, for the purpose of generation of initial conditions this simple choice is sufficient. 
Aside from the anisotropy coefficient $\beta_\mathrm{m}$, the other free parameter is the degree of 
rotation support $k$: the mean azimuthal streaming velocity is given by 
$\overline{v_\phi}=k\left(\overline{v_\phi^2}-\overline{v_R^2}\right)^{1/2}$. 
With this definition, $k=0$ corresponds to no net rotation and $k=1$ 
corresponds to equal velocity dispersions in $R$ and $\phi$ directions.

Another approach, which could more tailored to the particular properties of the potential, 
is based on the knowledge of periodic orbit families \citep[e.g.][]
{AthanassoulaBMP1983,ContopoulosGrosbol1989,Athanassoula1992,HasanPN1993,SkokosPA2002}. 
First, one constructs a ``periodic orbit map'', which consists of the initial conditions 
for various periodic orbit families (most importantly, prograde loop orbits) in the range of 
values of Jacobi constant $E_J$ from the bottom of the potential well to the corotation. 
Then one seeds initial conditions for the actual orbit library by picking them from the map 
and adding a random perturbation to the velocity components, to diversify the orbit library.
We have experimented with this approach, but did not find a noticeable improvement over 
the simpler method based on the axisymmetric Jeans equations; therefore, in the rest of the paper 
we use only the latter.

\subsection{Obtaining the self-consistent solution}  \label{sec:solution}

After computing the entire orbit library, the next step in the Schwarzschild method is 
to obtain the weights of orbits, such that the weighted superposition of density distributions 
of each orbit reproduces the target density profile of the model. Additionally, one often 
requires that some other constraints be satisfied, for instance, fitting the line-of-sight 
velocity profiles from the model to the observed values. The present implementation allows 
only for very rudimentary kind of kinematic constraints: one can specify the profile of 
velocity anisotropy coefficient $\beta$ as a function of radius. 
We will not use this feature for the models of discs, and will set $\beta=0$ for the haloes. 

The density profile of the model can be represented in a discretized way either as an array 
of masses of cells in a spatial grid, or, as suggested in \citet{Vasiliev2013}, as an array 
of coefficients describing the basis-set or spline spherical-harmonic expansion of the potential 
corresponding to this density. Since spherical-harmonic expansions are poorly suited for 
highly flattened systems, we use the former, traditional approach of partitioning 
the configuration space into a three-dimensional grid, computing masses of each grid cell 
$m_c$ by integrating the target density profile over the volume of the cell, and recording 
the fraction of time $t_{oc}$ that each orbit spends in each cell.
For elliptical galaxies, a grid based on concentric shells in radius and a certain partitioning 
in angles is typically used \citep[e.g.][]{MerrittFridman1996,vdBosch2008}.
For flattened systems we instead use a grid aligned with cylindrical coordinates (covering 
the region $0\le R\le R_\mathrm{max},0\le z\le z_\mathrm{max}$ with $N_R\times N_z$ rectangular 
cells, and split into $N_\phi$ cells in the azimuthal direction, equidistantly in $\phi$). 
The grid nodes in the meridional plane are assigned in such a way that the total mass in each slab 
($M_{R_k}\equiv\int_{R_k}^{R_{k+1}} dR \int_0^{2\pi} d\phi \int_{-\infty}^{\infty} dz\, R\rho(R,\phi,z)$ 
and a similar expression for $M_{z_k}$) is the same for each slice in $z$ direction or cylindrical 
shell in $R$ direction. Of course, if the density profile is not separable in $R,z$ then 
the masses of each cell are not equal, but they usually vary only by a factor of few.

The self-consistent solution for orbit weights $w_o$ is given by the matrix equation
$\sum_o w_o t_{oc} = m_c$, $c=1\dots N_\mathrm{cell}$. This is not just a standard linear-algebra 
problem, as we require the solution vector $w_o$ to be non-negative; in this formulation it is 
called a linear programming problem.
In practice, an exact solution may not always be feasible; moreover, it is worthwhile to apply 
some sort of non-linear regularization to improve the smoothness of the solution. 
A suitable formulation that meets these requirements is the quadratic programming problem:
minimize $\mathcal F\equiv \mathcal F_\mathrm{cell} + \mathcal F_\mathrm{orb}$
while keeping $\sum_{o=1}^{N_\mathrm{orb}} w_o t_{oc} - m_c = \delta_c$.
Here $\delta_c$ is the deviation in the mass of $c$-th cell from its required value. 
The penalty function $\mathcal F$ is split into two parts: one is responsible for 
minimizing the deviations from the self-consistent solution (which would have $\delta_c=0$), 
the other regularizes the orbit weights and optionally imposes additional penalties for using 
particular types of orbits (e.g.\ orbits with negative $z$-component of angular momentum, 
or chaotic orbits).

For the first task, several methods have been suggested in the past (here $\alpha_{\dots}$ 
are adjustable parameters): 
\begin{enumerate}
\item $\mathcal F_\mathrm{constr} = \alpha_1 \sum |\delta_c|$ \citep[e.g.][]{Siopis1998};
\item $\mathcal F_\mathrm{constr} = \alpha_2 \sum \delta_c^2$ \citep[least-square minimization, 
used in][]{MerrittFridman1996,Zhao1996b};
\item $\mathcal F_\mathrm{constr} = \sum \{ 0\mbox{ if }|\delta_c|<\alpha_0 m_c, \mbox{ otherwise }
\infty\}$ \citep[constraints should be satisfied within a given fractional error $\alpha_0$, as in]
[]{vdBosch2008}.
\end{enumerate}
The first method is easily implemented in the context of linear optimization problem by 
introducing $2N_\mathrm{cell}$ additional non-negative variables $\mu_c,\nu_c$ such that 
$\delta_c=\mu_c-\nu_c$; thus the penalty function is just proportional to the sum of these 
variables (of which at most one is non-zero for each constraint). 
Similarly, in the second method, which traditionally has been solved using 
the non-negative least-squares method \citep{LawsonHanson1974}, the introduction of 
the same additional variables transforms it to a quadratic optimization problem with 
$\mathcal F_\mathrm{constr} = \alpha_2 \sum_c(\mu_c^2+\nu_c^2)$, which we have found to be 
more numerically efficient. Finally, the last method can also be reformulated as a linear 
optimization problem with twice larger number of equations.
All these methods are available in \SMILE; we usually use a combination of first two methods, 
so that small deviations in cell masses are penalized linearly and large -- quadratically. 

The regularization method that we use for the second task just aims to achieve an uniform 
distribution of orbit weights:
\begin{align}  \nonumber
F_\mathrm{orb} = \frac{\lambda}{N_\mathrm{orb}} \sum_{o=1}^{N_\mathrm{orb}} (w_o/\tilde w_o)^2,
\end{align}
where $\tilde w_o$ is the weight prior (in the case of uniformly sampled initial conditions 
it is just the mean weight of an orbit in the model), and $\lambda$ is the regularization 
coefficient. Adjusting the values of $\alpha_{1,2}$ and $\lambda$, one may vary the relative 
severity of constraint violation versus irregularities in the distribution of orbit weights. 
In practice, for a well-behaved model with a sufficiently large number of orbits 
($N_\mathrm{orb} \gg N_\mathrm{cell}$), we expect all constraints to be satisfied exactly, 
thus only the second term in $\mathcal F$ remains nonzero. However, as we have discovered 
while working with potential approximations constructed from \Nbody snapshots, sometimes 
the mass of several cells could be computed from the density model with a rather large error
and cannot be satisfied by the Schwarzschild solution. In this case it may be better 
to sacrifice a few cells in order to keep the model reasonably smooth.

An extension over the previous version of \SMILE is the full support for multicomponent models. 
This is implemented as follows: the orbits are computed in a combination of several potential 
models (e.g.\ triaxial Ferrers bar plus an exponential disc plus a spherical Hernquist halo), 
and the solution of the self-consistent problem uses an arbitrary subset of these density models 
(e.g. the first two for the ``disc'' component and the last for the ``halo'' component, which 
are constructed independently and may use different methods for seeding the initial conditions).
While multicomponent Schwarzschild models have been used before \citep{CapuzzoLMV2007}, 
the present implementation is much more general and versatile. 
This extension required some modifications in the Jeans and Eddington initial condition samplers, 
to distinguish between the total potential of the entire model and the intrinsic density profile 
of a particular component.

\section{Tests}  \label{sec:tests}

\subsection{Axisymmetric disc--bulge--halo models}  \label{sec:axisymmetric}

In the first test, we construct composite models of an axisymmetric disc galaxy with a bulge 
and a halo, using several different methods:
\textsc{MaGalie} \citep{BoilyKP2001}, a descendant of Hernquist's (1993) \textsc{buildgal} code;
\textsc{mkgalaxy} \citep{McMillanDehnen2007};
\textsc{GalactICs} \citep{WidrowPD2008}, which is the latest version of the method of 
\citet{KuijkenDubinski1995};
\textsc{galic} \citep{YurinSpringel2014}; and \SMILE.
All these codes are publicly available: the first two as part of \textsc{nemo} toolbox
\citep{Teuben1995}, the third as part of \textsc{amuse} framework \citep{Pelupessy2013}, 
and the last two at the authors' websites.

\begin{figure} 
\includegraphics{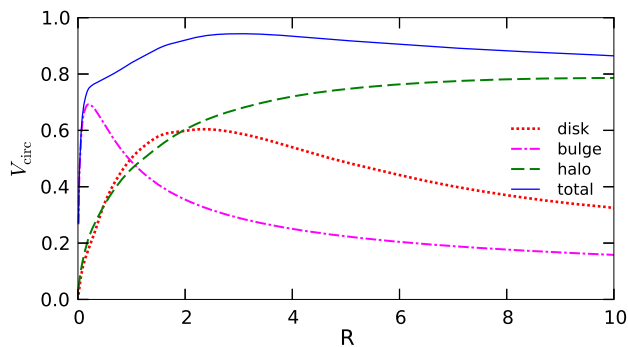}
\caption{
Circular velocity curve of the axisymmetric disc--bulge--halo model used in the test.
}  \label{fig:rotcurve_axi}
\end{figure} 

\begin{figure} 
\includegraphics{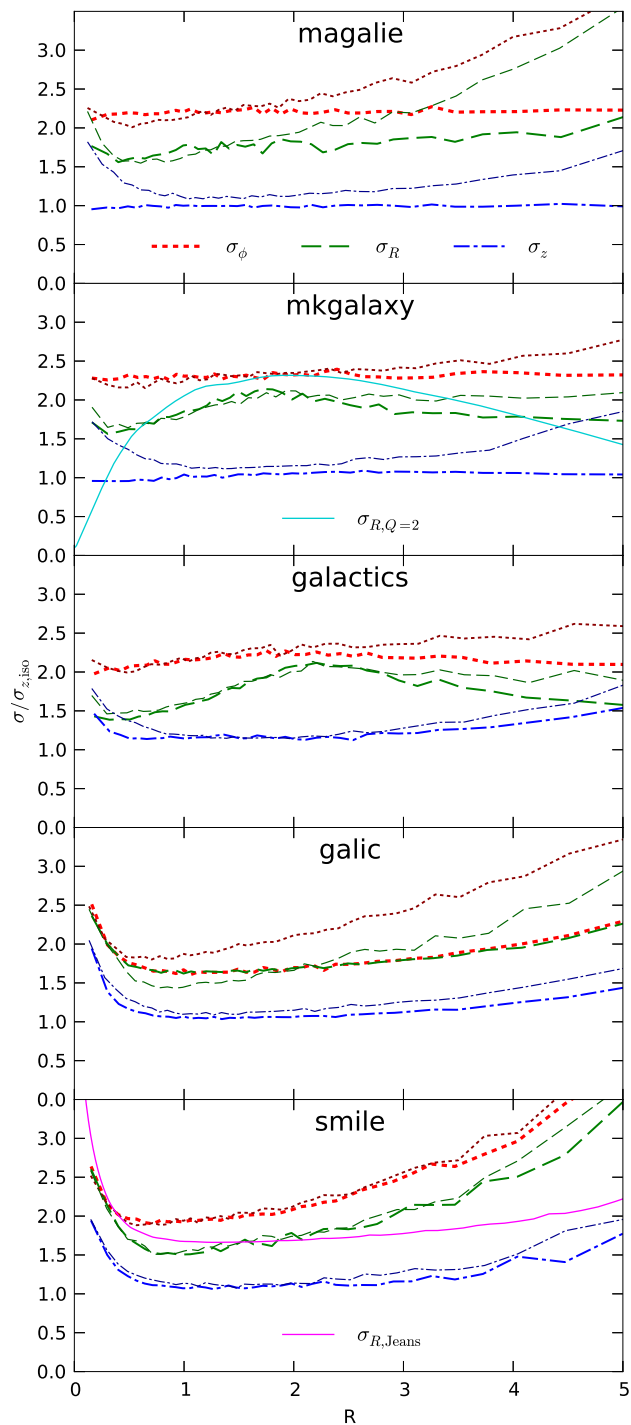}
\caption{
Velocity dispersion profiles of the disc component, for five different methods. Shown are values 
of $\sigma_R$ (dashed line), $\sigma_\phi$ (dotted line) and $\sigma_z$ (dot--dashed line), 
normalized by the value of vertical velocity dispersion for an isolated isothermal exponential disc:
$\sigma_{z,\mathrm{iso}}(R) \equiv \sqrt{G\,M_\mathrm{disc}\,z_0/(2R_0^2)}\,\exp(-R/(2R_0))$, 
where $M_\mathrm{disc}=1, R_0=1, z_0=1/8$ are the disc mass, scale radius and scaleheight.
\protect\\
Thicker lines are the initial models, thinner lines with shorter dashes are the models 
evolved for 100 time units. \protect\\
Also shown (in solid lines) are radial velocity profiles corresponding to the value of Toomre 
parameter $Q=2$ (in the second panel) and to the solution of axisymmetric Jeans equation 
with $\beta_\mathrm{m}=0.6$ (in the last panel, served as initial conditions for orbit library).
}  \label{fig:disk_kinem_axi}
\end{figure} 

The first complication that we encountered is the different specifications of galaxy parameters 
in these codes. All methods employ various parametrization of three components -- disc, bulge 
and halo density profiles, with fixed but sometimes incompatible functional forms, 
and in several cases, specified only indirectly and non-intuitively:
for instance, instead of component masses \textsc{galactics} uses the depth of gravitational 
potential, and the disc scalelength in \textsc{galic} is determined by its mass and an obscure 
halo spin factor.
After some experimentation, we have selected the following parameters, which produced nearly
identical mass models for all codes:
an exponential disc with density profile $\rho_\mathrm{disc} = M_\mathrm{disc}/(4\pi R_0^2 z_0)\;
\exp(-R/R_0)\;\mathrm{sech}^2(z/z_0)$; 
a bulge with scale radius $r_\mathrm{bulge}$ and either S\'ersic profile with the index $n=2$, 
or Hernquist profile; and a Navarro--Frenk--White (NFW) halo with cutoff at 10 scalelengths, 
or an equivalent Hernquist halo.
The masses and sizes of these components are: $M_\mathrm{disc}=1$, $R_0=1$, $z_0=1/8$, 
$M_\mathrm{bulge}=0.25$ (for the S\'ersic profile), $r_\mathrm{bulge}=0.2$, 
$M_\mathrm{halo} \approx 25$ (depending on the functional form of cutoff), and halo scale radius
of either 5 (for NFW) or 10 (for Hernquist). 
The rotation curve of this model is shown on Fig.~\ref{fig:rotcurve_axi}. 
The choice of disc thickness implies a moderately warm disc, and the masses and scale radii of 
spheroidal components give them a rather large contribution to the total gravity: both factors 
aim at avoiding disc instabilities, but at the same time stress the necessity to take into 
account the gravity of all mass components when constructing the disc distribution function.

Exponential disc is the standard option for all these codes, but the density profile of
the spheroidal components is constructed differently: \textsc{mkgalaxy} and \textsc{galactics} 
imply that the density profile follows the equipotential surfaces, which are somewhat flattened 
in the central parts. Consequently, the $z/x$ axis ratio of the bulge is close to 0.8 in these 
codes, so we adopted this value for the other methods. The halo axis ratio varies from 
$\sim 0.7$ in the centre to about 0.8 at $R=2$ (the maximum of disc rotation curve) and then 
approaches unity as $R$ increases further. For \SMILE we have incorporated this spatially variable 
flattening into the input density profile; the other two codes do not allow variable flattening 
and we have assumed a spherical halo for them. The velocity dispersion of bulge and halo was 
required to remain isotropic.
Most differences between codes are manifested in the velocity distribution of the disc component.

The models constructed with these five methods all had $(160,40,800)\times 10^3$ particles 
in the disc/bulge/halo components, and were checked for stability by evolving them for 
100 time units with the $N$-body code \textsc{gyrfalcon} \citep{Dehnen2000b,Dehnen2002}; 
the softening length was set to $0.01$ for disc and bulge and $0.03$ for halo particles, 
and the timestep was $2^{-7}$. 
Fig.~\ref{fig:disk_kinem_axi} shows the radial profiles of three diagonal components of 
velocity dispersion tensor in cylindrical coordinates -- thicker lines for the initially 
constructed models,and thinner lines for the evolved models. 
The velocities are rescaled in such a way that the vertical velocity dispersion of an isolated 
isothermal exponential disc would appear as a constant line equal to unity.

\begin{figure} 
\includegraphics{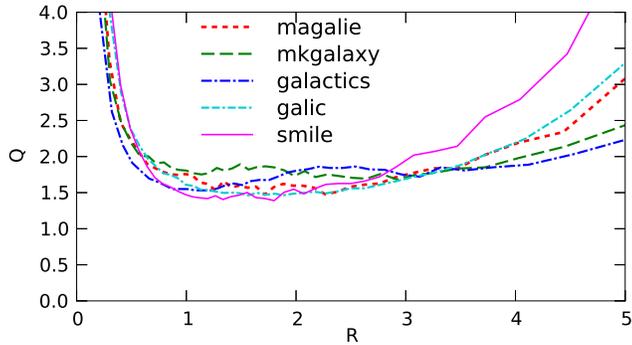}
\caption{
Toomre $Q$ parameter for the disc component of the axisymmetric models built with different methods.
}  \label{fig:disk_kinem_Q}
\end{figure} 

First thing to note is that the codes produce a diversity of velocity profiles, determined by 
different constraints. 
To assign the radial velocity dispersion, \textsc{magalie} and \textsc{mkgalaxy} 
use Toomre $Q$ parameter, \textsc{galactics} assumes its exponential dependence on radius, 
and \textsc{galic} and \SMILE solve axisymmetric Jeans equation. 
The vertical velocity dispersion is simply taken from an isolated isothermal disc 
in \textsc{magalie} and \textsc{mkgalaxy}, resulting in a too low value in the centre and 
in the outermost part of the disc (where there is a substantial contribution to the gravity 
from the bulge and the halo, correspondingly). \SMILE and \textsc{galic} take this contribution 
into account, while \textsc{galactics} is halfway between these options. 
\textsc{galic} enforces $\sigma_\phi=\sigma_R$, while other codes determine $\sigma_\phi$ from 
other considerations, resulting generally in $\sigma_\phi\gtrsim \sigma_R$. 
We note that the initial conditions for the orbit library in \SMILE are assigned with 
$\sigma_\phi=\sigma_R=\sigma_z/\sqrt{1-\beta_\mathrm{m}}$ (with the choice of 
$\beta_\mathrm{m}=0.6$), however the actual orbit integration results in a different velocity 
structure, which we do not constrain. Adding such constraints is not difficult, but as 
Figs~\ref{fig:disk_kinem_Q} and \ref{fig:disk_kinem_axi} (solid line in the bottom panel) show, 
neither a constant $Q$ nor a constant $\sigma_R/\sigma_z$ are good choices for the target 
velocity dispersion profile, thus raising the question of the most natural requirements for it.

The results of the simulations are also quite diverse. The models that assumed the vertical 
velocity dispersion to be equal to that of a self-gravitating isothermal disc (\textsc{magalie} and 
\textsc{mkgalaxy}) have quickly increased it in the central parts, while the disc thickness 
correspondingly dropped in the centre. 
In all models, the vertical thickness and velocity dispersion of the disc component has slowly 
but steadily grown over time, presumably as a result of two-body relaxation \citep{Sellwood2013}, 
and due to slowly developing disc instabilities.
\textsc{magalie} and \textsc{galic} did not preserve the initial profiles of $\sigma_R$ 
and $\sigma_\phi$ (the former code -- only in the outer parts). 
The biggest limitation of the latter code stems from enforcing $\sigma_R=\sigma_\phi$ 
(or, in other words, assuming the streaming parameter $k=1$): while there are other 
possible choices in the code, none corresponds to the actual variation of $k$ seen 
in other models. It might be better not to fix the value of $\sigma_\phi$ in this approach.
\textsc{galactics} generated a model reasonably close to equilibrium, and \SMILE fared best 
in this aspect. Note that in the latter model the velocity dispersion profile is considerably 
higher at large radii than in other methods. 
In terms of computer resources, \textsc{magalie} and \textsc{galactics} take a few minutes 
to create the models, \textsc{mkgalaxy} takes about an hour, \SMILE takes a few CPU hours 
(for the total number of $5\times10^4$ orbits in our models). \textsc{galic} took 
about 100 CPU hours for 100 iteration steps, but since the model taken after 10 iterations 
is essentially the same as after 100, a more truthful execution time
is about 10 CPU hours. \textsc{galic} used $\sim 32$~Gb of memory, while 
\SMILE needed about 1~Gb, and the other codes much less.

The results of this test demonstrate that \SMILE is capable of creating multicomponent systems 
in almost perfect equilibrium; a big improvement over previously used codes. 
It is important to note that, unlike other codes considered here, it allows an arbitrary 
density profile for (any number of) mass components, although in the present version 
it does not impose any constraints on the velocity structure. 

\subsection{Rotating triaxial Dehnen model}  \label{sec:dehnen}

As a second test, we attempt to construct models of mildly flattened, cuspy density profiles, 
typical for elliptical galaxies, having a varying degree of figure rotation. 
We consider a $\gamma=1$ \citet{Dehnen1993} model with axis ratios $x:y:z=1:0.75:0.5$, which 
rotates about its short axis. This is one of the models that was studied in \citet{DeibelVM2011} 
in the context of orbit analysis. 
The distinct feature of moderately cuspy ($\gamma=1$) triaxial Dehnen models is the rich 
network of resonant and thin%
\footnote{A thin orbit is confined to a two-dimensional, possibly self-intersecting sheet 
in the configuration space; its fundamental frequencies $\omega_i$ of motion in three directions 
satisfy a resonant relation $\sum_{i=1}^3 a_i\omega_i=0$ with integer $a_i$.}
orbits, which are clearly seen on frequency map plots (e.g.\ \citealt{ValluriMerritt1998}, fig.~9, 
or \citealt{VasilievAthanassoula2012}, fig.~2). 
Orbits associated with these resonances have a variety of shapes and are important ingredients 
of self-consistent models, replacing regular box orbits that appear in models with 
constant-density cores (the non-resonant box orbits are generally chaotic in cuspy potentials). 
The effect of rotation on regular box and thin orbits is the so-called ``envelope doubling''
\citep{deZeeuwMerritt1983}, caused by changing the sign of the Coriolis force as the orbit 
travels with positive or negative angular momentum. 
\citet{DeibelVM2011} have shown that this effect does not stabilize the chaotic orbits by 
converting them into regular ones that avoid passing through the centre, as has been 
suggested by \citet{GerhardBinney1985}: on the contrary, the fraction of chaotic orbits in 
rotating models is higher than in non-rotating ones \citep[cf.][]{Muzzio2006}, and increases 
with pattern speed, although some of them become converted to regular tube-like orbits when 
the rotation is high enough. 
Thus, the question raised by \citet{DeibelVM2011} is whether the self-consistent models 
could be constructed for moderate to high pattern speeds, and we address this question below.

\begin{figure} 
\includegraphics{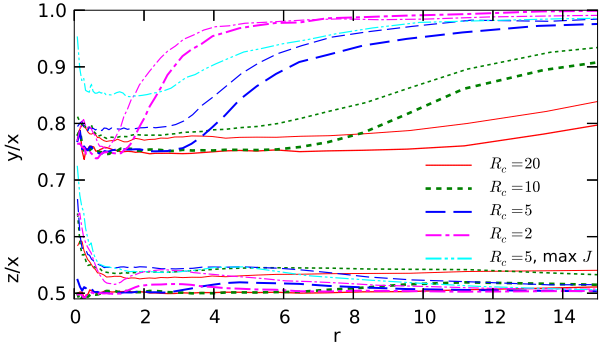}
\caption{
Axis ratios of four rotating Dehnen models considered in Section~\ref{sec:dehnen}, 
labelled by their corotation radius. Thick lines are the initial models, and thin lines show 
models evolved for 1000 time units. The original models had a variable $y:x$ axis ratio, 
ranging from 0.75 in the central part to 1 beyond the corotation radius. 
Evolved models largely maintained their initial shape, although becoming somewhat rounder 
(this effect was most noticeable for models with extreme angular momentum).
}  \label{fig:axis_ratio}
\end{figure} 

\begin{figure} 
\includegraphics{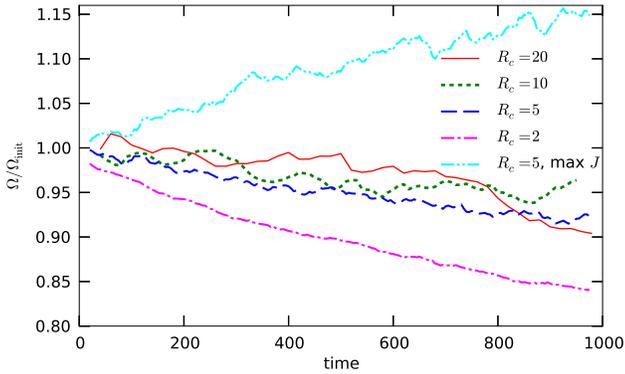}
\caption{
Evolution of pattern speed of four rotating Dehnen models, normalized to its initial value 
(which equals 0.0109, 0.0287, 0.0745, 0.257 for models with corotation radii 
$R_\mathrm{c}=20,10,5,2$, correspondingly).
}  \label{fig:patternspeed_dehnen}
\end{figure} 

\begin{figure} 
\includegraphics{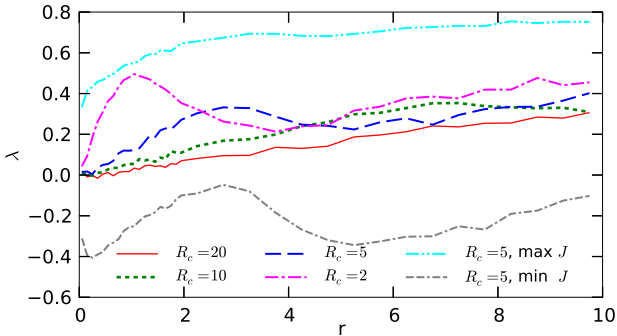}
\caption{
The rotation parameter $\lambda_\mathrm{R}$ \citep{Emsellem2007} for triaxial Dehnen models with 
figure rotation, seen along the intermediate ($y$) axis, plotted as a function of distance 
to the centre $r\equiv \sqrt{x^2+z^2}$; half of the mass lies within $r\approx 1.4$.
This parameter characterizes the amount of streaming motion, normalized to the total 
mean-square velocity; a value of 0.1 separates slow from fast rotators.
In the ``default'' models with no kinematic constraints, streaming motion is generally 
proportional to the pattern speed, but there is a lot of freedom in it: for two additional 
models with the same pattern speed but maximizing or minimizing the angular momentum, 
it varied in a much wider range (the latter actually was counter-rotating with the figure 
of potential, which is shown by a negative value of $\lambda_\mathrm{R}$).
}  \label{fig:angmom_lambda_dehnen}
\end{figure} 

We take several values for the pattern speed (see table~1 in that paper), corresponding to 
the corotation radius $R_\mathrm{c}$ equal to 20, 10, 5 and 2, and construct Schwarzschild models 
with $5\times10^4$ orbits, $\sim10^3$ constraints, and the initial conditions generated using 
either Eddington or axisymmetric Jeans equation (the latter produces more orbits that rotate 
in the same sense as the figure of potential, but the results did not differ significantly 
between these two approaches for these mildly flattened models).
It was quickly discovered that the model cannot maintain elongated shape beyond corotation, 
so we modified the density profile to have a variable $x:y$ axis ratio, which was kept at 
the value 0.75 up to $0.6R_\mathrm{c}$, and further out gradually changed towards unity 
(Fig.~\ref{fig:axis_ratio}); furthermore, we did not attempt to constrain the azimuthal 
distribution of mass outside $R_\mathrm{c}$.
We confirmed that the degree of chaoticity of all major orbit families in general increased 
with the pattern speed. Considering the orbit population in the region $R<2$, which encloses 
roughly half of total mass, we found that all models except the most rapidly rotating one
contain around 40\% short-axis tubes and 20\% resonant and thin orbits, while in the most
rapidly rotating model nearly 80\% orbits are short-axis tubes.

We then evolved \Nbody representations of each model for 1000 time units with \textsc{gyrfalcon},
using $5\times10^5$ particles and a softening length of 0.01. 
We measured the shape and orientation of the models, using both Fourier analysis of the surface 
density and the method of inertia tensor \citep[see][for a discussion of its variants]{Zemp2011}. 
The models turned out to be rather stable:
by the end of the simulation all of them largely retained their shape and pattern speed 
(Fig.~\ref{fig:patternspeed_dehnen}), the latter changed by at most 15\% (recall that 
the most rapidly rotating model has performed around 40 full revolutions by that time). 

We also briefly explored how the pattern speed is related to the angular momentum of our models. 
With the default settings, the total angular momentum turned out to be roughly proportional 
to the pattern speed. For the model with $R_\mathrm{c}=5$ we created two additional solutions that 
minimized and maximized the total angular momentum, adding corresponding terms into the penalty 
function \citep[e.g.][]{Pfenniger1984b}. The former resulted in a model where the streaming 
velocity and pattern speed were directed in the opposite sense, while in the latter case almost 
all orbits rotated in the same direction (Fig.~\ref{fig:angmom_lambda_dehnen}).
Both of these models were confirmed to be reasonably stable in \Nbody simulations; 
the model with maximum streaming motion actually had increased the pattern speed by $\sim 15\%$.
The axis ratios of these models by the end of simulation were closer to unity than for the 
``default'' models, although they still retained distinctly triaxial shapes.

\subsection{Barred disc galaxy model}  \label{sec:barreddisc}

For the next test, we consider an analytic potential model with a Miyamoto--Nagai disc, 
a Ferrers bar and a Plummer bulge, with the same parameters as in \citet{SkokosPA2002}; 
similar models were previously explored in \citet{Pfenniger1984a,HasanPN1993,PatsisSA2002,
ManosAthanassoula2011}, etc., in the context of orbit analysis, while \citet{Pfenniger1984b} 
has constructed two-dimensional (in the equatorial plane) Schwarzschild models for 
a similar density profile (without the bulge component). The pattern
speed is chosen so that the corotation radius is just beyond the end of the bar.
We combined the density profiles of all three components into one model and did not require 
the orbits to fit each of them separately, as in Section~\ref{sec:axisymmetric}.
We considered several choices for the number of orbits (up to $10^5$) and density constraints 
(up to $2\times10^3$), as well as for the generation of initial conditions (using axisymmetric 
Jeans equation with different choices of $\beta_\mathrm{m}$, and launching orbits from 
the vicinity of stable periodic orbits). 

\begin{figure} 
\includegraphics{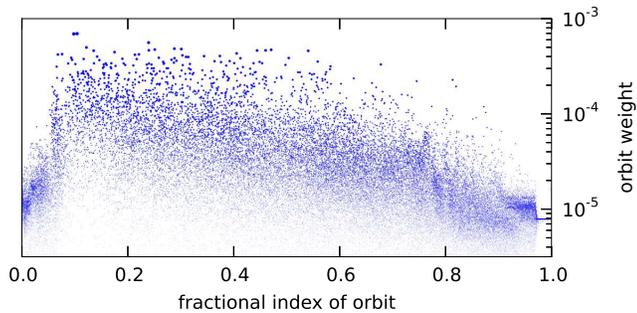}
\caption{
Distribution of orbit weights in a Schwarzschild model of a composite model from 
Section~\ref{sec:barreddisc}, as a function of orbit index (sorted in energy). 
In the ideal case, it should be narrowly spread around $1/N_\mathrm{orb}=10^{-5}$, 
but in this case this happens only at small or large energies; around the energy 
corresponding to the end of the bar the model barely manages to satisfy the constraints 
by using only a few orbits with relatively high weights. This usually indicates troubles 
in finding a self-consistent solution.
}  \label{fig:weights}
\end{figure} 

The results of all these experiments suggest that there is something unrealistic about 
the density model that we attempt to construct self-consistently. 
While the models with a coarse spatial grid could be constructed, the distribution of orbit 
weights in the solution turns out to be very uneven, especially in the region around the end 
of the bar (Fig.~\ref{fig:weights}).
Furthermore, a detailed inspection of the surface density map of the Schwarzschild model 
revealed that it did not exactly match the analytic model, having sharper edges at the end 
of the bar, even though their discretized versions were equal.
An attempt to refine the spatial grid resulted in the model becoming infeasible, no matter 
how many orbits we tried.
On the other hand, when we restricted the orbits to a two-dimensional plane and fitted 
the surface density, as in \citet{Pfenniger1984b}, the model was feasible and 
matched the projected density very well; however, with zero velocity in $z$ direction it 
is not in a dynamical equilibrium.

The model also turned out to be unstable when tested by an \Nbody simulation and quickly 
(within one rotation period) transformed itself into a rounder and more slowly-rotating shape, 
with a less prominent bar. 
We repeated the experiment with a model that lacked a central bulge, and it also 
demonstrated a similar slowdown of pattern speed (by some 30\% after five rotation periods),
while the bar became shorter and narrower. We thus rule out the
possibility that the bulge was the reason for this evolution.

\subsection{Reconstructing $N$-body snapshots of barred disc models from their density profiles}  
\label{sec:models_nbody}

In the previous sections, we used  analytically defined potential--density pairs to create 
our Schwarzschild models, while in this section we will use densities and potentials extracted 
from \Nbody simulations. The latter approach is, of course, much more realistic than the former, 
because the components are fully self-consistent instead of rigid, and, furthermore, because 
they have grown naturally under the influence of the total gravity of their system. 
Thus their shapes and mass distributions can be much more complex than what simple analytic 
formulae can describe. 

As simulations have shown, barred galaxies are not steady-state systems
\citep[for a review see][and references therein]{Athanassoula2013}. 
The bar region is ready to emit angular momentum, while the resonant regions in the outer 
disc and particularly in the halo are ready to absorb it \citep{LyndenBellKalnajs1972, 
TremaineWeinberg1984, Athanassoula2002, Athanassoula2003} so that there is a redistribution 
of angular momentum within the galaxy, as a result of which a number of the properties of 
the galaxy change \citep{LittleCarlberg1991a, LittleCarlberg1991b, DebattistaSellwood2000, 
AthanassoulaMisiriotis2002, Athanassoula2003, ONeillDubinski2003, ValenzuelaKlypin2003}. 
Thus bars become longer and stronger and their pattern speed decreases with time. 

Thus barred galaxies will necessarily have some non-stationarity and this goes against
the assumption of a steady state (at least in the rotating frame), implied by the Schwarzschild 
method. They therefore present an additional challenge for this method. Can such models be built? 
And, if yes, can their evolution match that of the density distribution we are trying to model, 
or at least provide some useful information about secular evolution?
These interesting questions will the subject of a future paper, so that only a short summary 
of the results relevant to the Schwarzschild modelling will be given here.

Although it is now well established that barred galaxies evolve, it is still unclear how strong 
this evolution is. Simulations have shown that the extent to which the angular momentum is 
redistributed, and therefore how important the corresponding changes in the galactic properties 
are, depends on the properties of the simulated galaxy. It is thus difficult, if at all possible, 
to use simulations to set constraints on evolution. Furthermore, observations show no consensus, 
since their results advocate evolution ranging from relatively little \citep{Perez2012} to quite 
important \cite[e.g.][]{KormendyKennicutt2004, ElmegreenEKBBP2007, Sheth2008, Cheung2013, Kim2015}. 
It is therefore necessary to consider both weak and strong secular evolution models.     

We will here attempt to model two examples of barred galaxies, extracted from simulations chosen 
so as to have as different secular evolutions as possible; one being very little (model A) and 
the other quite strong (model B). These two simulations are similar to the MD and MH ones 
considered by \citet{AthanassoulaMisiriotis2002}, having the same initial mass distributions 
for the halo and disc, i.e. the same rotation curves (see fig.~1 of that paper) and the same 
disc-to-halo mass ratio. Thus the halo of model B is much more centrally concentrated than that 
of model A. We take a snapshot from each of the two simulation at a time very near the beginning 
of its secular evolution phase and apply the Schwarzschild method to construct a steady-state, 
two-component dynamical model, using as constraints only the density distribution of both the 
disc and the halo. These are derived in a non-parametric way from the particle positions in 
the original snapshot. For the disc we also need to include figure rotation, i.e. we need an
estimate for the pattern speed. Since we want to test whether it is possible to build models 
in the absence of kinematic information, we cannot use the \citet{TremaineWeinberg1984} method 
to obtain the pattern speed $\Omega_\mathrm{b}$. Instead, we tried values in a range encompassing 
the ``correct'' one, i.e. the value obtained from the simulations.

The original \Nbody models had $N_\mathrm{disc}=2\times10^5$ disc particles and 
$N_\mathrm{halo}=10^6$ halo particles. 
In order to reduce the noise and transient features in the disc component, we followed 
\citet{Athanassoula2005} and \citet{IannuzziAthanassoula2015} and  
stacked five consecutive snapshots from the original simulation, after rotating them 
so as to align their bar major axes. The potential of the disc was represented with 
a cylindrical spline expansion with $m_\mathrm{max}=6$ azimuthal harmonics. 
For the halo we used a spline spherical-harmonic expansion, keeping only the monopole term 
for model A (which was found to be nearly-spherical), and additionally the quadrupole terms 
for model B (which is mildly triaxial in the central parts, see
\citealt{ColinVK2006, Athanassoula2007}).
We used $10^5$ orbits for each component, with the disc and halo density model discretized 
into $\sim1000$ and $\sim500$ cells, respectively.

The library of orbits for the halo component were created using the isotropic Eddington 
distribution function computed for a spherical approximation of the true density profile, 
while for the disc component we used the axisymmetric Jeans equations, with the meridional 
anisotropy parameter $\beta_\mathrm{m}=0.5$. A comparison of the latter with the kinematic 
properties of the disc in the original simulation snapshot shows that the assumption of constant 
$\beta_\mathrm{m}$ is clearly not very good: it varies from $\beta_\mathrm{m}\simeq 0.2$ 
in the centre to $\beta_\mathrm{m}\simeq 0.9$ at large radii.
Even so, we were able to construct Schwarzschild models for both cases A and B. 
These have a kinematic structure which is very similar to that of the original model, 
if the pattern speed is the ``correct'' one. Apparently, for these models there is not much 
freedom in the solution, so that even without any additional kinematic constraints (apart from 
adopting the ``correct'' pattern speed value) they turned out to be close to the original models. 

The feasibility of constructing a Schwarzschild model does not imply anything about its 
stability properties, or about the amount of secular evolution it will have once it is allowed 
to evolve. Note that this angular momentum exchange, and therefore evolution, can occur only 
if both the disc and the halo are represented as ``live'' particles, as opposed to a static 
potential. 

To test the evolution, we converted the Schwarzschild models to \Nbody representations using 
the same number of particles as the original snapshots. We then evolved them using two 
simulation codes: \textsc{gadget-2} \citep{Springel2005} and \textsc{gyrfalcon}; 
the results were quite similar \citep[cf.][]{FortinAL2011}, except that the \textsc{gadget} 
snapshots need to be re-centred. Below we use the data from the \textsc{gyrfalcon} simulations. 
The simulations were run for 800 time units (same units as in \citealt{AthanassoulaMisiriotis2002}); 
the original models performed about 12 full revolutions during this time.

\begin{figure} 
\includegraphics{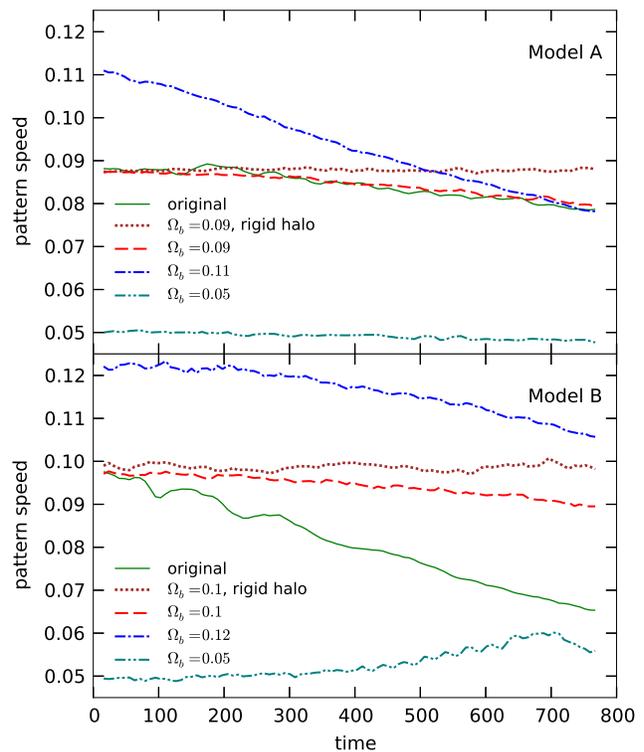}
\caption{
Pattern speed of model A (top) and model B (bottom) as a function of time. 
We show the curves corresponding to the original model (solid lines), 
and to Schwarzschild models with different choices for the pattern speed (dashed and 
dot--dashed lines); dotted line shows the model evolved in a fixed halo potential. 
}  \label{fig:pattern}
\end{figure} 

We first tested the internal stability of each disc component by performing simulations 
following the evolution of each disc  in the  static potential of its host rigid halo, 
represented by a spline-interpolated spherical-harmonic expansion. 
For this, we used the external potential feature in \textsc{gyrfalcon} with the potential 
computed using the \textsc{smilepot} library, essentially the same as in the original 
Schwarzschild model. The absence of any slowdown (Fig.~\ref{fig:pattern}, dotted lines) 
confirms that the models are indeed stable. 

The second test involves the full, two-component system. We first evolved the simulation 
snapshots and give the results for their pattern speed in Fig.~\ref{fig:pattern} (solid lines). 
That of model A did not substantially change over the course of the simulation, while that of 
model B decreased roughly by a third. 
We then evolved the Schwarzschild models and found a surprisingdiversity of behaviour. 
Models with pattern speeds higher than the ``correct'' value (i.e. the value obtained from 
the simulation) tend to slow down more rapidly, especially in the case A. 
Models with pattern speeds lower than ``correct'' evolved less, but in opposite directions for 
the two cases. Models with  the ``correct'' pattern speed evolved rather little. 
Thus, the evolution of model A is very similar to that of the snapshot it was built to represent. 
However, we found also little evolution for model B, i.e. this Schwarzschild model evolved
much less than the  snapshot it was built to represent. 
From \cite{Athanassoula2002}, we expect this difference to be linked to the properties of the halo 
rather than the disc. To test this, we created a ``hybrid'' model with the disc taken from 
the original simulation, and the halo created with the Schwarzschild method. This model showed 
little evolution, similar to the purely Schwarzschild model. We are examining the reasons for 
this behaviour in a separate study.

\section{Discussion}  \label{sec:discussion}

\subsection{Advantages and shortcomings of the Schwarzschild method}
\label{subsec:discuss-Sch}

In this paper, we have used the Schwarzschild orbit superposition method to construct 
several quite different self-consistent equilibrium galactic models.
This method is very general and powerful, but one has to keep in mind several considerations.
First, the basic assumption behind this and virtually every other method for constructing 
dynamical models is that the system under study is in a stationary state. 
This needs not be the case, for instance, if we model non-axisymmetric disc galaxies, 
in which a complex interplay between local and global instabilities, resonances, dynamical 
friction and other mechanisms leads to constant evolution, even in the absence of dissipation.
Diffusion of chaotic orbits may also play a role in driving the secular evolution of both 
elliptical \citep[e.g.][]{MerrittValluri1996,KandrupSiopis2003,VasilievAthanassoula2012} and 
disc galaxies \citep[e.g.][]{VoglisSK2006,ContopoulosHarsoula2013}.
Thus the models that we intend to be in a steady state, may turn out to evolve considerably, 
undermining our assumption. In application to real galaxies, it might be possible to create 
a model that matches all existing observations, but we cannot ensure that its evolution will 
be the same as for the original galaxy. The example of Section~\ref{sec:models_nbody}
shows that we could construct a nearly stationary model, while the original \Nbody simulation 
(not a real galaxy, though) continued to evolve substantially.

The latter example also reminds about another conceptual problem: we always use a certain  
set of constraints in our methods, but there might exist other possible factors that we do not 
take into account, but that may distinguish between otherwise similar models having different 
internal structure and dynamics. An obvious example is the non-uniqueness of deprojection 
of a non-spherical galaxy \citep{GerhardBinney1996}, 
or the insufficiency of using only the first two velocity moments to constrain the mass 
distribution, which could be alleviated by taking the full line-of-sight velocity distribution
\citep[e.g.][]{MerrittSaha1993,Gebhardt2000}.
In Section~\ref{sec:models_nbody}, we only used the information about particle positions, 
not velocities, to construct a model which turned out to be quite close to the original 
one in terms of velocity profiles; nevertheless, had we imposed additional kinematic 
constraints, the model could have different properties.
This also highlights another complication, specific to the Schwarzschild method. Namely, 
it provides a solution that satisfies the given constraints in the best possible way,
but such a solution may not be unique. In practice, this is circumvented by adding some 
regularization procedures \citep{RichstoneTremaine1988,CrettonZMR1999}, but a rigorously 
justified method for choosing one particular solution out of many is missing.

Yet another issue is related to the difficulties in interpreting the outcome of Schwarzschild 
modelling in terms of convergence, illustrated by the model of Section~\ref{sec:barreddisc}. 
Our practical experience suggests that for a well-behaved model, with a large margin of 
feasibility and the number of orbits exceeding the number of constraints by a large factor 
($\gtrsim 10$, i.e.\ an ``underconstrained'' model), 
regularization tends to create a narrowly peaked distribution of orbit weights. 
Conversely, if there are only a small number of orbits that get a high weight, this usually 
indicates that the required constraints are hard or impossible to satisfy exactly, implying 
that the chosen density profile might not be possible to realize with a self-consistent 
distribution function. 
Note, however, that a poor choice of initial conditions can also result in bad convergence, 
and that the inclusion of additional (e.g.\ kinematic) constraints may interfere with 
the ability of the solution to exactly match the discretized density profile; for this reason, 
the Schwarzschild code of \citet{vdBosch2008} tolerates a few percent relative error
in cell masses.

There are many different methods for constructing equilibrium models of disc galaxies, 
which we have reviewed in Section~\ref{sec:overview}, and tested some of them in 
Section~\ref{sec:axisymmetric}, but very few are suitable for non-axisymmetric (e.g.\ barred) 
galaxies, and no other implementation of Schwarzschild method is capable to deal with full 
three-dimensional models of multicomponent triaxial galaxies with figure rotation.
Although our code in the current version is not intended for creating models based on 
observational data, it nevertheless presents a first step towards this task. 
Most models of external galaxies and even the Milky Way are created using the approximation of 
axisymmetry, which may introduce substantial biases in recovered galaxy parameters, 
such as the mass-to-light ratio \citep{Thomas2007,Lablanche2012} and the mass of the central 
supermassive black hole \citep{vdBosch2010,Onken2014}.
Moreover, the assumption of constant coefficient $\beta_\mathrm{m}$ of velocity anisotropy in 
the meridional plane, commonly used in axisymmetric Jeans analysis \citep{Cappellari2008}, 
is clearly violated for most models that we have considered.
Thus having a more flexible modelling method will be increasingly important for the current 
and forthcoming large galactic surveys, such as \textsc{ATLAS}${}^\mathrm{3D}$ 
\citep{Cappellari2011} or \textsc{CALIFA} \citep{Sanchez2012}, that yield a wealth of 
kinematic data for a large sample of galaxies.
From the theoretical side, the interaction of non-axisymmetric discs with triaxial haloes 
presents a number of interesting effects \citep[e.g.][]{BerentzenSJ2006,MachadoAthanassoula2010}, 
thus a flexible Schwarzschild method for creating such composite models is a valuable asset.

\subsection{Applications and implications of our Schwarzschild models}
\label{subsec:applications}

\subsubsection{Axisymmetric disc--bulge--halo models}
\label{subsubsec:discuss-axisym}

Simulations have shown that galactic discs are prone to bar
instabilities \citep{Hohl1971}, with a growth rate which is function of many galaxy
properties, such as halo, bulge, or gas mass, velocity dispersion in
the disc etc. \citep[see][for a review]{Athanassoula2013}. Thus such
discs are not stationary. Yet in Section~\ref{sec:axisymmetric},
we showed that the \SMILE software is capable of constructing
axisymmetric multicomponent models of disc galaxies. How are these
two statements compatible?

It is interesting to note that, although we did not set any kinematic constraints, 
all the disc--bulge--halo models webuilt have high velocity dispersions in the disc, 
as can be seen in Fig.~\ref{fig:disk_kinem_Q}. For all models, the minimum $Q$ value 
is between 1.4 and 1.7, and occurs at radii between one and three disc scalelengths. 
Moreover, at radii less than one and more than three disc scalelengths, the $Q$ values 
are much larger (see Fig.~\ref{fig:disk_kinem_Q}). This ensures that the bar forms 
very slowly \citep{AthanassoulaSellwood1986, Athanassoula2003} and the model is 
sufficiently close to stationarity for a Schwarzschild model to be possible. 

The fact that, in the absence of kinematical constraints, the Schwarzschild method 
finds a relatively hot disc as a solution, underlines the necessity of such constraints 
when modelling real galaxies and also suggests that it may not be possible to model cold 
discs using this method, unless the bar instability is slowed down sufficiently by 
the halo, bulge or gas. This leaves a sufficiently large range of axisymmetric disc 
galaxies to which the Schwarzschild method can be applied, including the lenticulars, 
the early-type spirals, and in general any hot discs. 

\subsubsection{Rotating triaxial ellipsoids}
\label{subsubsec:discuss-triaxial}

In Subsection~\ref{sec:dehnen}, we built models for cuspy triaxial ellipsoids, which 
can be used to represent elliptical galaxies. While we have not attempted an extensive 
exploration of the relevant parameter space, our results indicate that, from the dynamical 
point of view, cuspy triaxial elliptical galaxies may exist in a broad range of pattern speeds.

Observational studies indicate that elliptical galaxies fall into two classes -- 
slow and fast rotators, with the former more likely to be triaxial \citep{Emsellem2007}.
In terms of the kinematic rotation parameter $\lambda_\mathrm{R}$ introduced in that paper, 
our models cover the whole range from slow to fast rotators (Fig.~\ref{fig:angmom_lambda_dehnen}). 
If we consider only models with no other constraint than the density distribution and the degree
of figure rotation, then we find a rough general trend in the same sense as observations, 
i.e. more triaxial galaxies have smaller $\lambda_\mathrm{R}$ values. 
If, however, we introduce further constraints, such as minimizing, or maximizing the global 
angular momentum, then the difference in the $\lambda_\mathrm{R}$ value is much bigger than 
that due to a change of the pattern speed.

Thus it becomes clear that further work on such models is necessary,
e.g. to extend the axisymmetric Schwarzschild models of
\cite{Cappellari2007} and answer whether it is possible to have
strongly triaxial configurations with considerable figure rotation and
large rotation parameters.

\subsubsection{Barred disc galaxy models}
\label{subsubsec:discuss-bars-analytic}

As a next step, we attempted to use the Schwarzschild method to construct models of 
galaxies composed of a disc, a halo and a strong bar component. The potential and density 
of each of these components were given by analytic functions, as in orbital structure studies. 
Although we made a large number of trials, we failed to construct a stable, self-consistent 
such model. More than one explanations can come to mind for this.

One possibility is that the simple superposition of a number of rigid components may be 
far from a reasonable dynamical system. Indeed, all these components interact gravitationally 
and that could substantially modify their density distributions. For example, it is known 
that discs do not stay axisymmetric in the presence of bars, and form spirals, rings, 
density minima/maxima around some Lagrangian points, etc., while haloes will not stay 
spherically symmetric at small radii if the potential of the thin discs dominates in
the inner parts of the galaxy. 

A second alternative, which to our eyes is the most likely one, is that, as discussed in 
\citet{AthanassoulaLSB2014} and \citet{Athanassoula2015}, the vertical mass distribution 
of Ferrers ellipsoids is unrealistic. The real shape of bars is much
more complex than a spheroid, 
having two components, one long and thin (both horizontally and vertically) and the other 
short and thick \citep{Athanassoula2005}. The latter is also known as the boxy or peanut bulge, 
or the barlens. Since the orbits in the bar potential are able reproduce this complex bar shape 
\citep[e.g.][]{Pfenniger1984a, PatsisSA2002}, they may not be able to reproduce the simple 
ellipsoidal shape of the Ferrers bars as well. A further argument in favour of this alternative 
is the fact that it has been possible to make two-dimensional models of barred galaxy potentials, 
as shown both by \citet{Pfenniger1984b} and in Section~\ref{sec:barreddisc}. Moreover, in 
Section~\ref{sec:models_nbody} we have been able to make a Schwarzschild model for barred 
galaxies having a realistic vertical distribution. 

\subsubsection{Barred $N$-body galaxy models}
\label{subsubsec:discuss-bars-Nbody}

We also attempted to build Schwarzschild models for two $N$-body
snapshots taken at the beginning of the secular evolution phase of the
bar and using only the densities as a constraint. One of the two
cases was specifically chosen from a simulation 
which showed hardly any bar evolution, while the other, on the
contrary, was chosen from a simulation with a very strong
evolution. For the former we were able to build a self-consistent
model, which, when evolved, followed the evolution of the snapshot.
For the latter, however, although we could build a self-consistent
model, its evolution was considerably less than that of the
simulation. This presumably means that, of all the velocity
distributions compatible with the density distribution in the
simulation, in absence of kinematical constraints, the Schwarzschild
model found the one which was nearest to stationarity (see also the
corresponding discussion in Section~\ref{subsubsec:discuss-axisym}).

This may set some limits to the applicability of the Schwarzschild
method to barred galaxies, if barred galaxies are found to have
non-negligible evolution. Indeed, as shown by \cite{Athanassoula2002,
  Athanassoula2003}, the resonances in the halo absorb a large fraction
of the angular momentum emitted from the bar region. Since no
kinematic constraints can be set on the haloes, the Schwarzschild
method may tend to provide self-consistent models fulfilling all
the density constraints as well as some kinematic constraints for the
disc, but whose evolution may be totally different from that of the
real galaxy. Thus more work is necessary here before the Schwarzschild
method can be massively applied to barred galaxies, including our own
Milky Way.
 
\section{Conclusions}  \label{sec:conclusions}

We have presented an implementation of the Schwarzschild orbit superposition method for 
creating equilibrium dynamical models of non-spherical stellar systems. 
This is a further development of the publicly available \SMILE code \citep{Vasiliev2013}, 
with the major new and improved features primarily targeting disc galaxies:
a new general-purpose potential approximation in terms of two-dimensional spline-interpolated 
Fourier expansion in cylindrical coordinates, which is accurate even for very flattened systems, 
full support for multicomponent systems and for figure rotation. 
The software is suitable not only for constructing self-consistent models with given properties, 
but also for analysing the orbital structure of a given potential, either specified analytically 
or taken from an \Nbody simulation \citep[e.g.][]{HarsoulaKalapotharakos2009, Athanassoula2012}. 
This opens up new possibilities for studying the properties of orbits relevant for the formation 
of bars, rings, spiral arms, etc., which traditionally have been explored by integrating the 
motion of test particles in a given analytic potential \citep{SkokosPA2002,RomeroGomezAAF2011,
MonariAH2013}, and could now be improved by taking these particles from a self-consistent model
while retaining the smooth potential for the orbit calculation. One may even use several 
snapshots from an \Nbody simulation to construct a time-dependent smooth potential approximation 
and use it for orbit integration \citep{ManosMachado2014}.
To facilitate these applications, the module for computing potential and forces is presented 
as a separate library \textsc{smilepot}, with \textsc{C} and \textsc{python} interfaces and 
bindings to other stellar-dynamical software tools: \textsc{nemo} \citep{Teuben1995}, 
\textsc{amuse} \citep{PortegiesZwart2013} and \textsc{galpy} \citep{Bovy2015}.

We have confirmed the possibility of constructing steady-state models of disc galaxies, both 
axisymmetric and barred, with the Schwarzschild method. 
We compared several existing codes for creating composite axisymmetric disc--bulge--halo models 
and found that \SMILE performs very well in this task, producing models that are closest to 
equilibrium. 
While elliptical galaxies are not the main subject of this paper, we also considered models 
resembling triaxial elliptical galaxies with density cusps and figure rotation. 
We have shown that they may exist in a wide range of pattern speeds and angular momenta, 
and are stable over many rotation periods. 
On the other hand, a barred disc model with an analytically defined potential, 
extensively used for orbit analysis in previous studies, turned out to be impossible to 
realize self-consistently, and the closest approximation to it was quite unstable dynamically.
Nevertheless, there exist models of barred disc galaxies embedded in nearly-spherical haloes, 
that appear to be in almost steady state. 
We have constructed two such models using density profiles extracted from \Nbody simulations, 
utilizing only information about particle positions. 
These models were shown to be reasonably stationary, and their kinematical properties were 
similar to the original \Nbody models if the pattern speed matched these original simulations. 
Interestingly, one of the original \Nbody models demonstrated considerable evolution (slowdown 
and growth of the bar), while the corresponding Schwarzschild model evolved much less.
We explore the reasons for this difference in a future work.

\textbf{Acknowledgements:}
EV is grateful for the hospitality of Laboratoire d'Astrophysique de Marseille during his visit,
thanks John Magorrian and other participants of the 2nd Gaia Challenge meeting 
for stimulating discussions, and acknowledges support from the National Aeronautics 
and Space Administration under grant no.\ NNX13AG92G.
EA acknowledges financial support to the DAGAL network from the People Programme 
(Marie Curie Actions) of the European Union's Seventh Framework Programme FP7/2007--2013/ 
under REA grant agreement number PITN-GA-2011-289313, from the CNES (Centre National d'Etudes 
Spatiales -- France) and from the PNCG (Programme National Cosmologie et Galaxies -- France).
EA also acknowledges HPC resources from GENCI--TGCC/CINES (Grants x2013047098 and x2014047098)
and from the Mesocentre of Aix--Marseille Universit\'e (program DIFOMER). 


\normalsize

\appendix

\section{Potential expansion in cylindrical coordinates}

To evaluate the potential at an arbitrary point in cylindrical coordinates $R,z,\phi$, 
we may use the following algorithm (see \citealt{BinneyTremaine}, section 2.6.2 for razor-thin 
discs and \citealt{Cuddeford1993} for axisymmetric discs with arbitrary vertical profile). 
Let the density be represented by a Fourier series
$$\rho(R,z,\phi) = \sum_{m=0}^{m_\mathrm{max}} \rho_m(R,z)\, \exp(im\phi),$$ 
so that 
$$\rho_m(R,z) = (2\pi)^{-1}\int_0^{2\pi} \rho(R,z,\phi)\exp(-im\phi).$$
The potential generated by this density is represented as 
$$\Phi(R,z,\phi) = \sum_{m=0}^{m_\mathrm{max}} \Phi_m(R,z)\, \exp(im\phi),$$
with each term in the expansion 
\begin{flalign*}
\Phi_m(R,z) =& -G \int_{-\infty}^{+\infty}\!\! dz'\! \int_0^{\infty}\!\! dR' \,
2\pi R'\,\rho_m(R',z')\, \Xi, \\
\Xi(R,z,R',z') \equiv& \int_0^\infty dk\, J_m(kR) J_m(kR')\, \exp(-k|z-z'|) .
\end{flalign*}
The last integral can be evaluated analytically \citep{GradshteynRyzhik,CohlTohline1999},
using the Legendre function of the second kind $Q$:
\begin{flalign*}
\Xi =& \frac{1}{\pi\sqrt{RR'}} Q_{m-1/2}\left( \frac{R^2+R'^2+(z-z')^2}{2RR'} \right) \; 
\mbox{if }R,R'>0, \\
\Xi =& (R^2+R'^2+(z-z')^2)^{-1/2}\,\delta_{m0}\;\mbox{if }R=0\mbox{ or }R'=0.
\end{flalign*}

For a discrete point mass set, the potential coefficients are computed as
$$\Phi_m(R,z) = -G \sum_{p=1}^N m_p\, \Xi(R,z,R_p,z_p)\, \exp im\phi_p.$$

The spline potential expansion in cylindrical coordinates is initialized by computing 
the coefficients $\Phi_m(R,z), m=0\dots m_\mathrm{max}$ either from a given analytic density profile, 
or from an \Nbody snapshot, on a two-dimensional grid in $R,z$ covering a rectangle 
$R\le R_\mathrm{max}, |z|\le z_\mathrm{max}$.
To improve resolution at small radii, we use logarithmic scaling of coordinates 
($R=\ln(1+R/R_0)$, and similar for $z$), and to reduce the approximation error at large radii, 
we scale the potential by a factor $\sqrt{R_0^2+R^2+z^2}$, so that for large $R,z$ 
the scaled coefficient for $m=0$ tends to a constant.
Here $R_0\equiv -GM_\mathrm{total}/\Phi_0(0,0)$ is a characteristic radial scale of the system. 
The evaluation of potential, forces and density at an arbitrary point within the box 
is done by computing interpolating splines or their derivatives. 
Outside the box, density is assumed to be zero, and the potential and forces are computed 
using a quadrupole approximation to the mass distribution.

\begin{figure*} 
\includegraphics{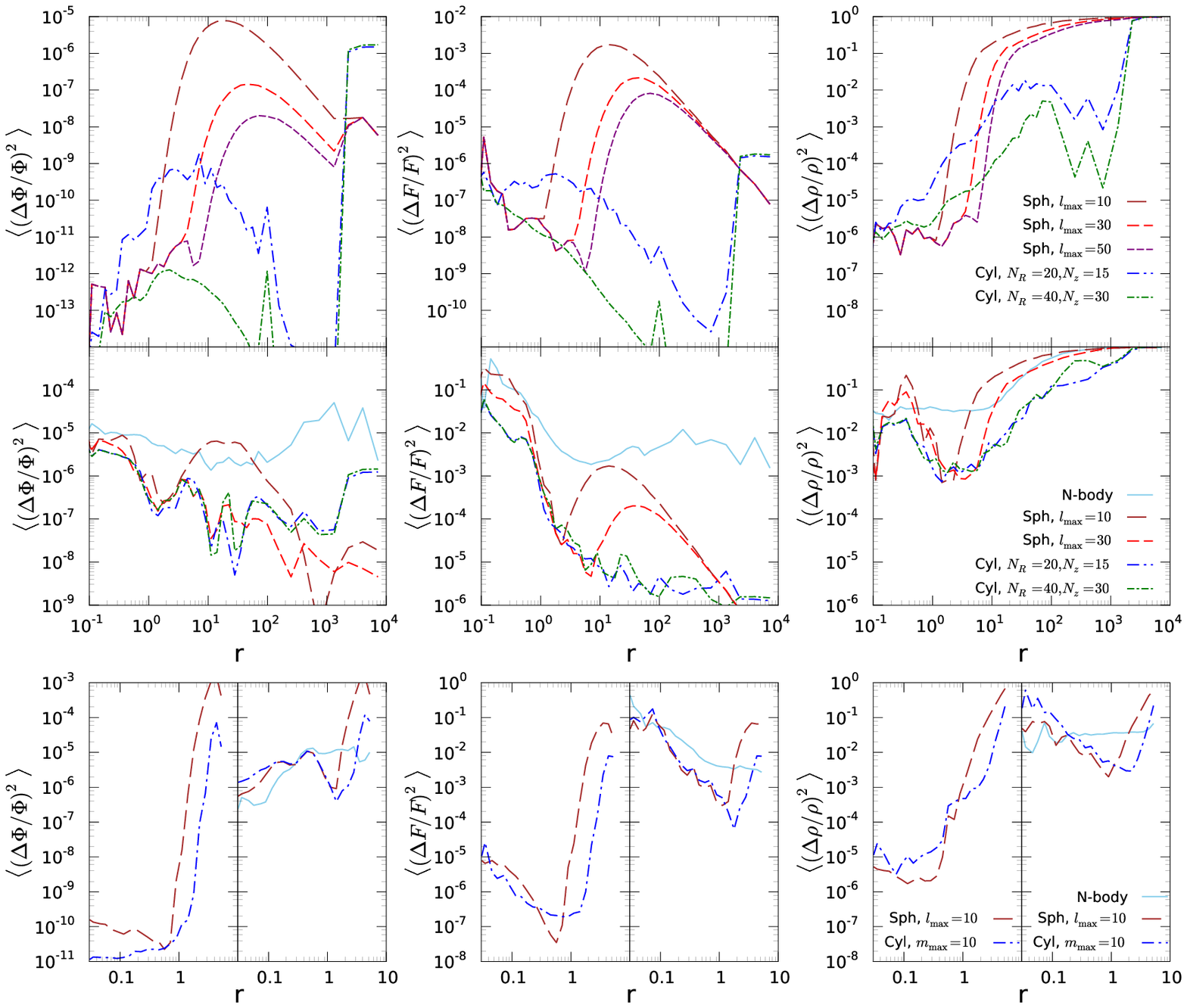}
\caption{
Mean square error in potential (left), force (centre), and density (right) approximation,
as functions of radius. \protect\\
\textbf{Top row:} axisymmetric Miyamoto--Nagai profile with $A=3,B=1$, using spherical-harmonic 
expansion (dashed lines) with the number of angular terms ranging from 10 to 50, 
and using spline interpolation on a grid in the meridional plane (dash--dotted lines), 
with different number of nodes (grid extends to $R=2000,z=100$).
Clearly, the latter provides much better accuracy than the spherical-harmonic expansion for 
the strongly flattened density profile, which converges very slowly. \protect\\
\textbf{Middle row:} the same density profile sampled by $N=10^5$ points.
In this case, the accuracy of cylindrical spline is limited by the discreteness noise at 
all radii (increasing the grid size does not make it smaller), 
while for the spherical-harmonic expansion the number of angular terms still limits 
the accuracy at large radii, where the density profile is very flattened. 
Note that both smooth potential expansions still approximate the target model better than 
a direct evaluation of potential and density from the \Nbody snapshot. \protect\\
\textbf{Bottom row:} strongly triaxial Ferrers model with axes $x:y:z=6:1.5:0.6$. 
Here, we compared the spherical-harmonic expansion with the radial basis set from spherical 
Bessel functions \citep[as in][]{AllenPP1990} with 40 radial terms, and the cylindrical 
spline with the grid size $N_R=20,N_z=15$; in both cases, the number of angular terms was 10.
Again the latter was found to approximate the target model better than the former.
Left half of each panel compares the approximations initialized from a smooth density profile, 
while the right half does the same for a $N=10^5$ particle snapshot; again in the latter case 
the discreteness noise limits the accuracy.
}  \label{fig:accuracy}
\end{figure*}

Fig.~\ref{fig:accuracy} demonstrates that for strongly flattened models, the cylindrical 
spline expansion is superior to the spherical-harmonic expansion, although if the potential 
is initialized from an \Nbody snapshot, the accuracy is mainly limited by discreteness noise.

\label{lastpage}

\begin{thebibliography}{}
\footnotesize

\bibitem[Allen, Palmer \& Papaloizou(1990)]{AllenPP1990}
Allen J., Palmer P., Papaloizou J., 1990, MNRAS, 242, 576

\bibitem[Athanassoula(1992)]{Athanassoula1992}  
Athanassoula E., 1992, MNNRAS, 259, 328

\bibitem[Athanassoula(2002)]{Athanassoula2002}  
Athanassoula E., 2002, MNRAS, 569, 83

\bibitem[Athanassoula(2003)]{Athanassoula2003}  
Athanassoula E., 2003, MNRAS, 341, 1179

\bibitem[Athanassoula(2005)]{Athanassoula2005}  
Athanassoula E., 2005, MNRAS 358, 1477

\bibitem[Athanassoula(2007)]{Athanassoula2007}  
Athanassoula E., 2007, MNRAS, 377, 1569

\bibitem[Athanassoula(2012)]{Athanassoula2012}  
Athanassoula E., 2012, MNRAS, 426, L46

\bibitem[Athanassoula(2013)]{Athanassoula2013}  
Athanassoula E., 2013, in J. Falc\'on-Barroso J. and Knapen J.H., eds,
\textsl{Secular Evolution of Galaxies}, Cambridge University Press, Cambridge, p. 305

\bibitem[Athanassoula(2015)]{Athanassoula2015}
Athanassoula E., 2015, in Laurikainen E., Peletier R., Gadotti D., eds.,
\textsl{Galactic Bulges}; Astrophysics and Space Science Library, vol.418, Springer (arXiv:1503.04804)

\bibitem[Athanassoula et al.(1983)]{AthanassoulaBMP1983}
Athanassoula E., Bienaym\'e O., Martinet L., Pfenniger D., 1983, A\&A, 127, 349

\bibitem[Athanassoula et al.(2014)]{AthanassoulaLSB2014}  
Athanassoula E., Laurikainen E., Salo H., Bosma A., 2014, MNRAS, submitted (arXiv:1405.6726)

\bibitem[Athanassoula \& Misiriotis(2002)]{AthanassoulaMisiriotis2002}
Athanassoula E., Misiriotis A., 2002, MNRAS, 330, 35

\bibitem[Athanassoula \& Sellwood(1986)]{AthanassoulaSellwood1986}
Athanassoula E., Sellwood J. A., 1986, MNRAS, 221, 213

\bibitem[Barnes(1988)]{Barnes1988}
Barnes J., 1988, ApJ, 331, 699

\bibitem[Berentzen et al.(1998)]{Berentzen1998}
Berentzen I., Heller C., Shlosman I., Fricke K., 1998, MNRAS, 300, 49

\bibitem[Berentzen, Shlosman \& Jogee(2006)]{BerentzenSJ2006}
Berentzen I., Shlosman I., Jogee S. ApJ, 637, 582

\bibitem[Binney \& McMillan(2011)]{BinneyMcMillan2011}  
Binney J., McMillan P., 2011, MNRAS, 413, 1889

\bibitem[Binney \& Tremaine(2008)]{BinneyTremaine}
Binney J., Tremaine S., 2008, \textsl{Galactic dynamics}, 2nd edn., Princeton Univ. Press, Princeton, NJ

\bibitem[Bissantz, Debattista \& Gerhard(2004)]{BissantzDG2004}  
Bissantz N., Debattista V., Gerhard O., 2004, ApJ, 601, L155

\bibitem[Boily, Kroupa \& Pe\~narrubia(2001)]{BoilyKP2001}  
Boily Ch., Kroupa P., Pe\~narrubia J., 2001, New Astron., 6, 27

\bibitem[Bovy(2015)]{Bovy2015}
Bovy J., 2015, ApJS, 216, 29

\bibitem[Brown \& Papaloizou(1998)]{BrownPapaloizou1998}
Brown M., Papaloizou J., 1998, MNRAS, 300, 135

\bibitem[Cappellari(2008)]{Cappellari2008}  
Cappellari M., 2008, MNRAS, 390, 71

\bibitem[Cappellari, Emsellem \& Bacon(2007)]{Cappellari2007}  
Cappellari M., Emsellem E., Bacon R., 2007, MNRAS, 379, 418

\bibitem[Cappellari, Emsellem \& Krajnovi\'c(2011)]{Cappellari2011}  
Cappellari M., Emsellem E., Krajnovi\'c D., 2011, MNRAS, 413, 813

\bibitem[Capuzzo-Dolcetta et al.(2007)]{CapuzzoLMV2007}
Capuzzo-Dolcetta R., Leccese L., Merritt D., Vicari A., 2007, ApJ, 666, 165

\bibitem[Cheung et al. (2013)]{Cheung2013}
Cheung, E. et al. 2013, ApJ, 779, 162

\bibitem[Clutton-Brock(1972)]{CluttonBrock1972}
Clutton-Brock M., 1972, Ap\&SS, 16, 101

\bibitem[Clutton-Brock(1973)]{CluttonBrock1973}
Clutton-Brock M., 1973, Ap\&SS, 23, 55

\bibitem[Cohl \& Tohline(1999)]{CohlTohline1999}
Cohl H., Tohline J., 1999, ApJ, 527, 86

\bibitem[Col\'\i n, Valenzuela \& Klypin(2006)]{ColinVK2006}
Col\'\i n P., Valenzuela O. \& Klypin A., 2006, ApJ, 644, 687

\bibitem[Contopoulos \& Grosb{\o}l(1988)]{ContopoulosGrosbol1988}
Contopoulos G., Grosb{\o}l P., 1988, A\&A, 197, 83

\bibitem[Contopoulos \& Grosb{\o}l(1989)]{ContopoulosGrosbol1989}
Contopoulos G., Grosb{\o}l P., 1989, A\&AR, 1, 261

\bibitem[Contopoulos \& Harsoula(2013)]{ContopoulosHarsoula2013}  
Contopoulos G., Harsoula M., 2013, MNRAS, 436, 1201

\bibitem[Contopoulos \& Papayannopoulos(1980)]{ContopoulosP1980}
Contopoulos G., Papayannopoulos T., 1980, A\&A, 92, 33

\bibitem[Cretton et al.(1999)]{CrettonZMR1999} 
Cretton N., de Zeeuw T., van der Marel R., Rix H.-W., 1999, ApJS, 124, 383

\bibitem[Cuddeford(1993)]{Cuddeford1993}
Cuddeford P., 1993, MNRAS, 262, 1076

\bibitem[Debattista \& Sellwood(2000)]{DebattistaSellwood2000}
Debattista V., Sellwood J., 2000, ApJ, 543, 704

\bibitem[Deibel, Valluri \& Merritt(2011)]{DeibelVM2011}
Deibel A., Valluri M., Merritt D., 2011, ApJ, 728, 128

\bibitem[Dehnen(1993)]{Dehnen1993} 
Dehnen W., 1993, MNRAS, 265, 250

\bibitem[Dehnen(2000)a]{Dehnen2000a}
Dehnen W., 2000, AJ, 119, 800

\bibitem[Dehnen(2000)b]{Dehnen2000b}
Dehnen W., 2000, ApJ, 536, L39

\bibitem[Dehnen(2002)]{Dehnen2002}
Dehnen W., 2002, J.Comp.Phys., 179, 27

\bibitem[Dehnen(2009)]{Dehnen2009}  
Dehnen W., 2009, MNRAS, 395, 1079

\bibitem[Dehnen \& Binney(1998)]{DehnenBinney1998}
Dehnen W., Binney J., 1998, MNRAS, 294, 429

\bibitem[de Lorenzi et al.(2007)]{deLorenziDGS2007}
de Lorenzi F., Debattista V., Gerhard O., Sambhus N., 2007, MNRAS, 376, 71

\bibitem[de Zeeuw \& Merritt(1983)]{deZeeuwMerritt1983}
de Zeeuw T., Merritt D., 1983, ApJ, 267, 571

\bibitem[Earn(1996)]{Earn1996}
Earn D., 1996, ApJ, 465, 91

\bibitem[Elmegreen et al.(2007)]{ElmegreenEKBBP2007}
Elmegreen B., Elmegreen D., Knapen J., Buta R., Block D., Puerari I., 2007, ApJ, 670, 97

\bibitem[Emsellem et al.(2007)]{Emsellem2007}
Emsellem E., Cappellari M., Krajnovi\'c D., et al., 2007, MNRAS, 379, 401

\bibitem[Fortin, Athanassoula \& Lambert(2011)]{FortinAL2011}  
Fortin P., Athanassoula E., Lambert J.-C., 2011, A\&A, 531, 120

\bibitem[Fragkoudi et al.(2015)]{Fragkoudi2015}
Fragkoudi F., Athanassoula E., Bosma A., Iannuzzi F., 2015, MNRAS, 450, 229

\bibitem[Fux(2001)]{Fux2001}
Fux R., 2001, A\&A, 373, 511

\bibitem[Gebhardt et al.(2000)]{Gebhardt2000}
Gebhardt K., Richstone D., Kormendy J., et al., 2000, AJ, 119, 1157

\bibitem[Gerhard \& Binney(1985)]{GerhardBinney1985}
Gerhard O., Binney J., 1985, MNRAS, 216, 467

\bibitem[Gerhard \& Binney(1996)]{GerhardBinney1996}
Gerhard O., Binney J., 1996, MNRAS, 279, 993

\bibitem[Gradshteyn \& Ryzhik(1965)]{GradshteynRyzhik}
Gradshteyn I., Ryzhik I., 1965, \textsl{Tables of integrals, series and products}, 4th edn., 
Academic Press, New York

\bibitem[H\"afner et al.(2000)]{HafnerEDB2000}  
H\"afner R., Evans N., Dehnen W., Binney J., 2000, MNRAS, 314, 433

\bibitem[Harsoula \& Kalapotharakos(2009)]{HarsoulaKalapotharakos2009}
Harsoula M., Kalapotharakos C., 2009, MNRAS, 394, 1605

\bibitem[Hasan, Pfenniger \& Norman(1993)]{HasanPN1993}
Hasan H., Pfenniger D., Norman C., 1993, ApJ, 409, 91

\bibitem[Hernquist(1993)]{Hernquist1993}  
Hernquist L., 1993, ApJS, 86, 389

\bibitem[Hernquist \& Ostriker(1992)]{HernquistOstriker1992}
Hernquist L., Ostriker J., 1992, ApJ, 386, 375

\bibitem[Hohl(1971)]{Hohl1971}  
Hohl F. 1971, ApJ, 168, 343

\bibitem[Holley-Bockelmann, Weinberg \& Katz(2005)]{HolleyWK2005}
Holley-Bockelmann K., Weinberg M., Katz N., 2005, MNRAS, 363, 991

\bibitem[Hunt \& Kawata(2013)]{HuntKawata2013}  
Hunt J., Kawata D., 2013, MNRAS, 430, 1928

\bibitem[Hunt, Kawata \& Martel(2013)]{HuntKM2013}  
Hunt J., Kawata D., Martel H., 2013, MNRAS, 432, 3062

\bibitem[Iannuzzi \& Athanassoula(2015)]{IannuzziAthanassoula2015}
Iannuzzi F., Athanassoula E., 2015, MNRAS, 450, 2514

\bibitem[Kalapotharakos, Patsis \& Grosb\o l(2010)]{KalapotharakosPG2010}
Kalapotharakos C., Patsis P., Grosb{\o}l P., 2010, MNRAS, 408, 9

\bibitem[Kandrup \& Siopis(2003)]{KandrupSiopis2003}
Kandrup H., Siopis C., 2003, MNRAS, 345, 727

\bibitem[Kaufmann \& Contopoulos(1996)]{KaufmannContopoulos1996}
Kaufmann D., Contopoulos G., 1996, A\&A, 309, 381

\bibitem[Kazantzidis, Magorrian \& Moore(2004)]{KazantzidisMM2004} 
Kazantzidis S., Magorrian J., Moore B., 2004, ApJ, 601, 37

\bibitem[Kim et al.(2015)]{Kim2015} 
Kim T., Sheth K., Gadotti D., et al., 2015, ApJ, 799, 99

\bibitem[Kormendy \& Kennicutt(2004)]{KormendyKennicutt2004} 
Kormendy J., Kennicutt R., 2004, ARA\&A, 42, 603 

\bibitem[Kuijken \& Dubinski(1995)]{KuijkenDubinski1995}
Kuijken K., Dubinski J., 1995, MNRAS, 277, 1341

\bibitem[Lablanche et al.(2012)]{Lablanche2012}
Lablanche P.-Y., Cappellari M., Emsellem E., et al., 2012, MNRAS, 424, 1495

\bibitem[Lawson \& Hanson(1974)]{LawsonHanson1974}
Lawson C., Hanson R., 1974, \textsl{Solving least-squares problems}, Prentice--Hall, NJ

\bibitem[Little \& Carlberg(1991a)]{LittleCarlberg1991a}  
Little B., Carlberg R., 1991a, MNRAS, 250, 161

\bibitem[Little \& Carlberg(1991b)]{LittleCarlberg1991b}  
Little B., Carlberg R., 1991b, MNRAS, 251, 227

\bibitem[Long \& Mao(2010)]{LongMao2010}  
Long R., Mao S., 2010, MNRAS, 405, 301

\bibitem[Long et al.(2013)]{LongMSW2013}  
Long R., Mao S., Shen J., Wang Y., 2013, MNRAS, 428, 3478

\bibitem[Lynden-Bell \& Kalnajs(1972)]{LyndenBellKalnajs1972}  
Lynden-Bell D., Kalnajs A. J., 1972, MNRAS, 157, 1

\bibitem[McGlynn(1984)]{McGlynn1984}
McGlynn T., 1984, ApJ, 281, 13

\bibitem[Machado \& Athanassoula(2010)]{MachadoAthanassoula2010}
Machado R., Athanassoula E., 2010, MNRAS, 406, 2386

\bibitem[McMillan \& Dehnen(2007)]{McMillanDehnen2007}
McMillan P., Dehnen W., 2007, MNRAS, 378, 541

\bibitem[Manos \& Athanassoula(2011)]{ManosAthanassoula2011}
Manos T., Athanassoula E., 2011, MNRAS, 415, 629

\bibitem[Manos \& Machado(2014)]{ManosMachado2014}
Manos T., Machado R., 2014, MNRAS, 438, 2201

\bibitem[Merritt \& Fridman(1996)]{MerrittFridman1996} 
Merritt D., Fridman T., 1996, ApJ, 460, 136

\bibitem[Merritt \& Saha(1993)]{MerrittSaha1993} 
Merritt D., Saha P., 1993, ApJ, 409, 75

\bibitem[Merritt \& Valluri(1996)]{MerrittValluri1996} 
Merritt D., Valluri M., 1996, ApJ, 471, 82

\bibitem[Monari, Antoja \& Helmi(2013)]{MonariAH2013}
Monari G., Antoja T., Helmi A., 2013, arXiv:1306.2632

\bibitem[Muzzio(2006)]{Muzzio2006} 
Muzzio J., 2006, Celest.\ Mech.\ Dyn.\ Astron., 96, 85

\bibitem[O'Neill \& Dubinski(2003)]{ONeillDubinski2003} 
O'Neill J.~K., Dubinski J., 2003, MNRAS, 346, 251

\bibitem[Onken et al.(2014)]{Onken2014}  
Onken C., Valluri M., Brown J., et al., 2014, ApJ, 791, 37

\bibitem[Patsis, Skokos \& Athanassoula(2002)]{PatsisSA2002}
Patsis P., Skokos Ch., Athanassoula E., 2002, MNRAS, 337, 578

\bibitem[P{\'e}rez, Aguerri \& M{\'e}ndez-Abreu(2012)]{Perez2012} 
P{\'e}rez I., Aguerri J.~A.~L., M{\'e}ndez-Abreu J., 2012, A\&A, 540, A103 

\bibitem[Pelupessy et al.(2013)]{Pelupessy2013}
Pelupessy I., van Elteren A., de Vries N., McMillan S., Drost N.,
Portegies Zwart S. 2013, A\&A, 557, A84

\bibitem[Pfenniger(1984a)]{Pfenniger1984a}  
Pfenniger D., 1984a, A\&A, 134, 373

\bibitem[Pfenniger(1984b)]{Pfenniger1984b}  
Pfenniger D., 1984b, A\&A, 141, 171

\bibitem[Pfenniger \& Friedli(1991)]{PfennigerFriedli1991}  
Pfenniger D., Friedli D., 1991, A\&A, 252, 75

\bibitem[Pfenniger \& Friedli(1993)]{PfennigerFriedli1993}  
Pfenniger D., Friedli D., 1993, A\&A, 270, 561

\bibitem[Portegies Zwart et al.(2013)]{PortegiesZwart2013}
Portegies Zwart S., McMillan S., van Elteren E., Pelupessy I., de Vries N., 2013, 
Comput.\ Phys.\ Commun., 184, 3, 456

\bibitem[Qian(1993)]{Qian1993}
Qian E., 1993, MNRAS, 263, 394

\bibitem[Richstone \& Tremaine(1988)]{RichstoneTremaine1988}  
Richstone D., Tremaine S., 1988, ApJ, 327, 82

\bibitem[Robin et al.(2003)]{RobinRDP2003}  
Robin A., Reyl\'e C., Derri\`ere S., Picaud S., 2003, A\&A, 409, 523

\bibitem[Robijn \& Earn(1996)]{RobijnEarn1996}
Robijn F., Earn D., 1996, MNRAS, 282, 1129

\bibitem[Rodionov, Athanassoula \& Sotnikova(2009)]{RodionovAS2009}
Rodionov S., Athanassoula E., Sotnikova N., 2009, MNRAS, 392, 904

\bibitem[Romero-G\'omez et al.(2006)]{RomeroGomez2006}
Romero-G\'omez M., Masdemont J., Athanassoula E., Garc\'\i a-G\'omez C., 2006, A\&A, 453, 39

\bibitem[Romero-G\'omez et al.(2011)]{RomeroGomezAAF2011}
Romero-G\'omez M., Athanassoula E., Antoja T., Figueras F., 2011, MNRAS, 418, 1176

\bibitem[Sanchez et al.(2012)]{Sanchez2012}  
S\'anchez S., Kennicutt R., Gil de Paz A., et al., 2012, A\&A, 538, 8

\bibitem[Schwarzschild(1979)]{Schwarzschild1979}
Schwarzschild M., 1979, ApJ, 232, 236

\bibitem[Schwarzschild(1982)]{Schwarzschild1982}  
Schwarzschild M., 1982, ApJ, 263, 599

\bibitem[Schwarzschild(1993)]{Schwarzschild1993}  
Schwarzschild M., 1993, ApJ, 409, 563

\bibitem[Sellwood(2003)]{Sellwood2003}  
Sellwood J., 2003, ApJ, 587, 638

\bibitem[Sellwood(2013)]{Sellwood2013}  
Sellwood J., 2013, ApJ, 769, L24

\bibitem[Sellwood \& Valluri(1997)]{SellwoodValluri1997}  
Sellwood J., Valluri M., 1997, MNRAS, 287, 124

\bibitem[Sheth et al.(2008)]{Sheth2008}
Sheth K., Elmegreen D., Elmegreen B., et al., 2008, ApJ, 675, 1141

\bibitem[Siopis(1998)]{Siopis1998}
Siopis C., 1998, \textsl{Nonuniqueness and Structural Stability of 
Self-consistent Models of Elliptical Galaxies}, PhD thesis, Univ. Florida

\bibitem[Skokos, Patsis \& Athanassoula(2002)]{SkokosPA2002}
Skokos Ch., Patsis P., Athanassoula E., 2002, MNRAS, 333, 847

\bibitem[Springel(2005)]{Springel2005}
Springel V., 2005, MNRAS, 364, 1105

\bibitem[Syer \& Tremaine(1996)]{SyerTremaine1996}
Syer D., Tremaine S., 1996, MNRAS, 282, 223

\bibitem[Teuben(1995)]{Teuben1995}
Teuben P., 1995, in Shaw R. A., Payne H. E., Hayes J. J. E., eds, ASP Conf. Ser. Vol. 77, 
Astronomical Data Analysis Software and Systems IV. Astron. Soc. Pac., San Francisco, p. 398

\bibitem[Teuben \& Sanders(1985)]{TeubenSanders1985}
Teuben P., Sanders R., 1985, MNRAS, 212, 257

\bibitem[Thakur et al.(2007)]{Thakur2007} 
Thakur P., Jiang I.-G., Das M., Chakraborty D.K., Ann H.B., 2007, A\&A, 475, 821

\bibitem[Thomas et al.(2007)]{Thomas2007}
Thomas J., Jesseit R., Naab T., Saglia R., Burkert A., Bender R., 2007, MNRAS, 381, 1672

\bibitem[Tremaine \& Weinberg(1984)]{TremaineWeinberg1984}
Tremaine S., Weinberg M., 1984, ApJ, 282, L5

\bibitem[Valluri \& Merritt(1998)]{ValluriMerritt1998}
Valluri M., Merritt D., 1998, ApJ, 506, 686

\bibitem[Valluri, Merritt \& Emsellem(2004)]{ValluriME2004}
Valluri M., Merritt D., Emsellem E., 2004, ApJ, 602, 66

\bibitem[van den Bosch et al.(2008)]{vdBosch2008} 
van den Bosch R., van de Ven G., Verolme E., Cappellari M., de Zeeuw T., 2008, MNRAS, 385, 647

\bibitem[van den Bosch \& de Zeeuw(2010)]{vdBosch2010}
van den Bosch R., de Zeeuw T., 2010, MNRAS, 401, 1770

\bibitem[Valenzuela \& Klypin(2003)]{ValenzuelaKlypin2003}
Valenzuela O., Klypin A., 2003, MNRAS, 345, 406 

\bibitem[Vasiliev(2013)]{Vasiliev2013}
Vasiliev, E. 2013, MNRAS, 434, 3174

\bibitem[Vasiliev \& Athanassoula(2012)]{VasilievAthanassoula2012} 
Vasiliev E., Athanassoula E., 2012, MNRAS, 419, 3268

\bibitem[Voglis, Stavropoulos \& Kalapotharakos(2006)]{VoglisSK2006}  
Voglis N., Stavropoulos I., Kalapotharakos C., 2006, MNRAS, 372, 901

\bibitem[Wang et al.(2012)]{WangZMR2012}  
Wang Y., Zhao H., Mao S., Rich R., 2012, MNRAS, 427, 1429

\bibitem[Wang et al.(2013)]{WangMLS2013}  
Wang Y., Mao S., Long R., Shen J., 2013, MNRAS, 435, 3437

\bibitem[Weinberg(1999)]{Weinberg1999} 
Weinberg M., 1999, AJ, 117, 629

\bibitem[Widrow \& Dubinski(2005)]{WidrowDubinski2005}
Widrow L., Dubinski J., 2005, ApJ, 631, 838

\bibitem[Widrow, Pym \& Dubinski(2008)]{WidrowPD2008}
Widrow L., Pym B., Dubinski J., 2008, ApJ, 679, 1239

\bibitem[Wozniak \& Pfenniger(1997)]{WozniakPfenniger1997}
Wozniak H., Pfenniger D., 1997, A\&A, 317, 14

\bibitem[Yurin \& Springel(2014)]{YurinSpringel2014}
Yurin D., Springel V., 2014, MNRAS, 444, 62

\bibitem[Zemp et al.(2011)]{Zemp2011}
Zemp M., Gnedin O., Gnedin N., Kravtsov A. 2011, ApJS, 197, 30

\bibitem[Zhao(1996a)]{Zhao1996a}  
Zhao H.-S., 1996a, MNRAS, 278, 488 

\bibitem[Zhao(1996b)]{Zhao1996b}  
Zhao H.-S., 1996b, MNRAS, 283, 149 

\end{thebibliography}
\end{document}